\documentclass[superscriptaddress,amsmath, nofootinbib]{revtex4}
\usepackage{amsfonts}
\usepackage{graphicx}
\usepackage[brazil]{babel}
\usepackage[latin1]{inputenc}
\usepackage{amsmath}
\usepackage{amssymb}
\begin{document}


\title{Lagrangian and Hamiltonian formulations of asymmetric rigid body, considered as a constrained system.}

\author{Alexei A. Deriglazov }
\email{alexei.deriglazov@ufjf.br} \affiliation{Depto. de Matem\'atica, ICE, Universidade Federal de Juiz de Fora,
MG, Brazil} 

\author{}

\date{\today}

\begin{abstract}
This work is devoted to a systematic exposition of the dynamics of a rigid body, considered as  a system with kinematic constraints. Having accepted the variational problem in accordance with this, we no longer need any additional postulates or assumptions about the behavior of the rigid body. All the basic quantities and characteristics of a rigid body, as well as the equations of motion and integrals of motion, are obtained from the variational problem by direct and unequivocal calculations within the framework of standard methods of classical mechanics. Several equivalent forms for the equations of motion of rotational degrees of freedom are deduced and discussed on this basis. Using the resulting formulation, we revise some cases of integrability, and discuss a number of peculiar properties, that are not always taken into account when formulating the laws of motion of a rigid body. 
\end{abstract}

\maketitle 

\tableofcontents


\section{Introduction.}

A rigid body can be defined as a system of $n$ particles, the distances and angles between which do not change with time. From point of view of classical mechanics,  we are dealing with a system subject to kinematic (that is velocity independent) constraints. Our task will be to formulate and discuss the dynamics of rigid body on this ground.   
Due to the constraints imposed on  $3n$ coordinates of the particles, only six coordinates turn out to be independent.  
The theory of a rigid body, including the convenient equations of motion for the independent degrees of freedom, was formulated by 
Euler, Lagrange and Poisson already at the dawn of the development mechanics \cite{Eul_1758, Lag_1788, Poi_1842}, and enters now as a chapter in the standard books on classical mechanics \cite{Whit_1917, Mac_1936, Lei_1965, Landau_8, Gol_2000, Arn_1, Arn_2, Grei_2003}. However, a didactically systematic formulation of the equations of motion is regarded not an easy task \cite{1,2,3,4,5,6,7}. For instance, J. E. Marsden, D. D. Holm and T. S. Ratiu in their work \cite{Mar_98} dated by 1998 write: "It was already clear in the last century that certain mechanical systems resist the usual canonical formalism, either Hamiltonian or Lagrangian, outlined in the first paragraph. The rigid body provides an elementary example of this."

Given mechanical system, its equations of motion can either be postulated or derived from a suitable variational problem. Here we adopted the second possibility. Let us point out some advantages of this approach. First, the formulation of a variational problem for a system with kinematic constraints is a well-understood problem in mechanics, and is used already when formulating the simplest systems such as a mathematical pendulum.  In a very short summary, it works as follows. Consider a mechanical system with generalised coordinates $q^A(t)$ and the Lagrangian $L(q^A, \dot q^A)$. Suppose the "particle" $q^A$ was then forced to move on a surface given by the algebraic 
equations $\chi_\alpha(q^A)=0$. Then equations of motion  is known to follow from the action functional, where the constraints are taken into account with help of auxiliary variables $\lambda_\alpha(t)$  as follows \cite{Arn_1, deriglazov2010classical}:  
\begin{eqnarray}\label{0.1}
S=\int dt ~~ L(q^A, \dot q^A)-\sum_\alpha \lambda_\alpha\chi_\alpha(q^A). 
\end{eqnarray}
The auxiliary dynamical variables $\lambda_\alpha(t)$ are called Lagrangian multipliers. In all calculations they should be treated on equal footing with $q^A(t)$. In particular, looking for the equations of motion, we take independent variations with respect to $q^A$ and  
all $\lambda_\alpha$.
The variation with respect to $\lambda_\alpha$, implies $\chi_\alpha(q^A)=0$, that is the constraints arise as a part of conditions of extremum of the action functional. So the presence of $\lambda_\alpha$ allows $q^A$ to be treated as independent variables, that should be varied independently in obtaining the equations of motion.  

Second, the  formalism for constructing the equations of motion from this variational problem also is  well-known. It is the Dirac's version of the Hamiltonian formalism, that works even for a more general (velocity dependent) constraints \cite{Dir_1950, GT, deriglazov2010classical}. While the constraints are taken into account with use of auxiliary variables, the formalism allows to remove all them  from final equations. Moreover, for any system with kinematic constraints, one can even write out closed expressions for the equations of motion that no longer contain the auxiliary variables, see Sect. 8.6.1  in \cite{deriglazov2010classical}. 

Third, we will not need to introduce in advance the basic quantities necessary to formulate the theory (vector of angular velocity, inertia tensor and so on). They are defined and discussed in the place where they directly arise in the study of a rigid body. This makes clearer both the reasonableness of their introduction and their physical and mathematical meaning. 

Fourth, we will be able to postpone the use of particular parameterizations like the Euler angles up to the place where they are really useful.  It should be noted that in the presentations based on the Euler angles, some of equations of motion (namely the Poisson equations (\ref{6.8})) are so dissolved in the calculations that sometimes even not mentioned. 

In this work we follow the above methodology for the case of a rigid body, considered as a system with constraints. It will be shown that all basic quantities, equations of motion and integrals of motion follow from this formalism in a systematic and natural way. This can be compared with the standard approach \cite{Mac_1936,Lei_1965,Gol_2000,Landau_8}, where a number of postulates should be assumed: on the behavior of the center of mass, as well as on  the conservation of energy and angular momentum.
Although this work is mainly of a pedagogical nature, in the Conclusion we list a number of specific properties of the theory of a rigid body, which are not always taken into account in the literature, when formulating the laws of motion and applying them.

{\bf Notation.} Capital letters of the Latin alphabet $N, P, A, B, \ldots$ or Greek latters $\alpha, \beta, \ldots$ are used to label particles. Latin letters $i, j, k, \ldots$ used to label coordinates. Vectors are denoted using the bold letters, for instance the position vector of the particle $N$ is ${\bf y}_N=(y_N^1, y_N^2, y_N^3)$, where $y_N^i$ are Cartesian coordinates of the particle. 

Summation over particles is always explicitly stated: $\sum_{N=1}^{n}m_N{\bf y}_N$. Repeated latin indices are summed unless otherwise indicated: $\epsilon_{ijk}y_N^j y_P^k=\sum_j \sum_k \epsilon_{ijk}y_N^j y_P^k$.  

Notation for the scalar product:  $({\bf a}, {\bf b})=a_i b_i$. Notation for the vector product: $[{\bf a}, {\bf b}]_i=\epsilon_{ijk}a_j b_k$, 
where $\epsilon_{ijk}$ is Levi-Chivita symbol in three dimensions, with $\epsilon_{123}=+1$. 

Recall that the sets of three-vectors and antisymmetric $3\times 3$ matrices are equivalent. For the vector $\boldsymbol\omega$, the corresponding matrix is denoted by $\hat\omega$, and we have the relationship
\begin{eqnarray}\label{0.1.1}
\boldsymbol\omega=\left(
\begin{array}{c}
\omega_1 \\
\omega_2 \\
\omega_3 
\end{array}\right) \quad \leftrightarrow \quad \hat\omega=\left(
\begin{array}{ccc}
0 & \omega_3 & -\omega_2 \\
-\omega_3 & 0 & \omega_1 \\
\omega_2 & -\omega_1 & 0  
\end{array}\right). 
\end{eqnarray}
For the components, we get 
\begin{eqnarray}\label{0.2}
\hat\omega_{ij}=\epsilon_{ijk}\omega_k, \qquad \omega_k=\frac12 \epsilon_{kij}\hat\omega_{ij}.  
\end{eqnarray}
From the definition $\det B=\frac{1}{6}\epsilon_{ijk}B_{ia}B_{jb}B_{kc}\epsilon_{abc}$, we have the useful identity 
\begin{eqnarray}\label{0.3}
\epsilon_{abc}=(\det B)^{-1}\epsilon_{ijk}B_{ia}B_{jb}B_{kc}. 
\end{eqnarray}

\section{Initial variational problem, translational and rotational degrees of freedom.}

Consider a system of $n\ge 4$ particles with the position vectors ${\bf y}_N(t)=(y_{N}^1(t), y_{N}^2(t), y_{N}^3(t))$ and masses $m_N$, $N=1, 2, \ldots , n$, not all lying on the same plane. The system is called a rigid body, if distances and angles between the particles do not depend of time 
\begin{eqnarray}\label{1.1}
({\bf y}_N(t)-{\bf y}_K(t),  {\bf y}_P(t)-{\bf y}_M(t))=\mbox{const}.  
\end{eqnarray}
The task is to write the equations of motion, that determine all functions ${\bf y}_N(t)$, if the initial positions and velocities of the particles are known. Denote the initial positions ${\bf y}_N(0)={\bf c}_N$, where $c_N^i$ are $3n$ given numbers, and by ${\bf v}_N$ the initial velocities of the particles.  

{\bf The number of independent degrees of freedom of the rigid body.}  Some of the constraints (\ref{1.1}) are consequences of others. We separate an independent subset of $3n-6$ of them and show, how this can be used to represent all vector functions ${\bf y}_N(t)$ through some six functions, that are no longer limited by the constraints. Hence to describe the dynamics of the rigid body, we only  need  to know the temporal evolution for these six functions. The rigid body is said to have six independent degrees of freedom. 

\begin{figure}[t] \centering
\includegraphics[width=08cm]{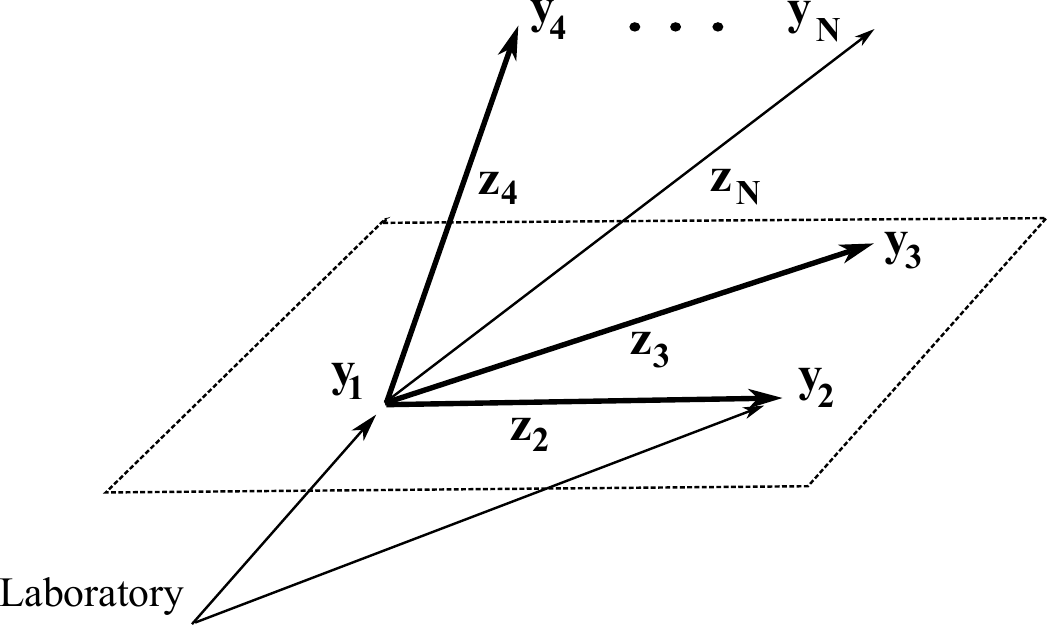}
\caption{Linearly independent vectors ${\bf z}_2(t), {\bf z}_3(t)$ and ${\bf z}_4(t)$ connecting four points of a body.}\label{TvT_2}
\end{figure}

Let's pick four points ${\bf y}_1, {\bf y}_2, {\bf y}_3$ and ${\bf y}_4$ not lying on the same plane. Then the vectors ${\bf z}_2={\bf y}_2-{\bf y}_1$, ${\bf z}_3={\bf y}_3-{\bf y}_1$ and ${\bf z}_4={\bf y}_4-{\bf y}_1$ are linearly independent, see Figure \ref{TvT_2}.  
Let us introduce the set of $n$ vectors ${\bf y}_1$, ${\bf z}_N={\bf y}_N-{\bf y}_1$, $N=2, 3, \ldots n$.  The constraints (\ref{1.1}) then read 
\begin{eqnarray}\label{1.1.1}
({\bf z}_N, {\bf z}_P)=\mbox{const}, \qquad 
({\bf z}_N-{\bf z}_K, {\bf z}_P-{\bf z}_M)=\mbox{const}.    
\end{eqnarray}
Consider the  subset of all constraints that contain ${\bf z}_2, {\bf z}_3$ and ${\bf z}_4$ 
\begin{eqnarray}
({\bf z}_A, {\bf z}_B)=a_{AB}, \qquad A, B=2, 3, 4,  \quad {\bf z}_A ~\mbox{are linearly independent vectors,  so} ~\det a\ne 0,  \label{1.2.1}\\
({\bf z}_A, {\bf z}_\alpha)=a_{A\alpha}, \qquad \alpha=5, 6, \ldots , n.  \label{1.2.2}\qquad \qquad \qquad \qquad  \qquad \qquad \qquad \qquad \qquad \qquad \qquad \quad 
\end{eqnarray}
Eqs.  (\ref{1.2.1}) fix the lengths of three vectors ${\bf z}_A$ and angles between them ($6$ constraints), while Eqs. (\ref{1.2.1}) fix  scalar products among each of $n-4$ vectors ${\bf z}_\alpha$  with three ${\bf z}_A$ ($3(n-4)$ constraints). So we have separated $3(n-4)+6=3n-6$ constraints. By construction, the remaining constraints of the set (\ref{1.1.1})  do not imply any restrictions on ${\bf z}_A$. Besides, there are no constraints on the vector ${\bf y}_1$. Let us write ${\bf z}_\alpha$ in the basis ${\bf z}_A$
\begin{eqnarray}\label{1.3}
{\bf z}_\alpha=k_\alpha^2 {\bf z}_2+k_\alpha^3 {\bf z}_3+k_\alpha^4 {\bf z}_4, 
\end{eqnarray}
and take the scalar products of this expression with ${\bf z}_2, {\bf z}_3$, and ${\bf z}_4$. The resulting system of equations allows to determine the coordinates $k_\alpha^A$ through the numbers $a_{AN}$: $k_\alpha^A=a^{-1}_{AB}a_{B\alpha}$. By this, all the vectors ${\bf y}_1, {\bf y}_2, \ldots, {\bf y}_n$ are represented through ${\bf y}_1, {\bf z}_2, {\bf z}_3$ and ${\bf z}_4$ as follows
\begin{eqnarray}\label{1.4}
{\bf y}_2={\bf y}_1+{\bf z}_2, \quad {\bf y}_3={\bf y}_1+{\bf z}_3, \quad {\bf y}_4={\bf y}_1+{\bf z}_4, \quad 
{\bf y}_\alpha={\bf y}_1+k_\alpha^A{\bf z}_A.  
\end{eqnarray}
Further, there are 6 restrictions (\ref{1.2.1}) on 9 coordinates of vectors ${\bf z}_A$, this gives $9-6=3$ independent degrees of freedom. For instance, they can be the Euler angles, that fix the position of the rigid triple ${\bf z}_A$ with respect to the coordinate axes of the laboratory. Three more independent degrees of freedom are the coordinates of the vector ${\bf y}_1$. 

{\bf Initial variational problem.}  Let us write the Lagrangian variational problem for the rigid body. To this aim, we write the independent constraints (\ref{1.2.1}) and (\ref{1.2.2}) in terms of initial variables as follows: 
\begin{eqnarray}
({\bf y}_A-{\bf y}_1, {\bf y}_B-{\bf y}_1)=a_{AB}, \qquad A, B=2, 3, 4,  \label{1.5.1}\\
({\bf y}_A-{\bf y}_1, {\bf y}_\alpha-{\bf y}_1)=a_{A\alpha}, \qquad  \alpha=5, 6, \ldots , n.  \label{1.5.2} 
\end{eqnarray}
They define a six-dimensional surface in ${\mathbb R}^ {3n}$, so the rigid body is represented by a point in ${\mathbb R}^ {3n}$, freely moving on this surface. According to the results known from classical mechanics \cite{Arn_1, deriglazov2010classical}, the Lagrangian action of this system is 
\begin{eqnarray}\label{1.6}
S=\int dt ~\frac12\sum_{N=1}^{n}m_N\dot{\bf y}_N^2+\frac12\sum_{A=2}^{4}\sum_{N=2}^{n}\lambda_{AN}\left[({\bf y}_A-{\bf y}_1, {\bf y}_N-{\bf y}_1)-a_{AN}\right]. 
\end{eqnarray}
The first term is kinetic energy of all particles, while the second term accounts the presence of the constraints. 
The auxiliary dynamical variables $\lambda_{AN}(t)$ are called Lagrangian multipliers. In all calculations they should be treated on equal footing with ${\bf y}_N(t)$. In particular, looking for the equations of motion, we take variations with respect to ${\bf y}_N$ and  all $\lambda_{AN}$. The $3\times3$\,-block $\lambda_{AB}$ of $\lambda_{AN}$ was chosen to be the symmetric matrix. The variations with respect 
to $\lambda_{AN}$ imply the constraints (\ref{1.5.1}) and (\ref{1.5.2}), while the variations with respect to ${\bf y}_N(t)$ give the dynamical equations
\begin{eqnarray}\label{1.7}
m_1\ddot{\bf y}_1=-\sum_{AB}\lambda_{AB}[{\bf y}_B-{\bf y}_1]-\frac12\sum_{A\alpha}\lambda_{A\alpha}[{\bf y_A}+{\bf y}_\alpha-2{\bf y}_1], \cr
m_A\ddot{\bf y}_A=\sum_{B}\lambda_{AB}[{\bf y}_B-{\bf y}_1]+\frac12\sum_{\alpha}\lambda_{A\alpha}[{\bf y}_\alpha-{\bf y}_1], \cr
m_\alpha\ddot{\bf y}_\alpha=\frac12\sum_{A}\lambda_{A\alpha}[{\bf y}_A-{\bf y}_1].
\end{eqnarray}

{\bf The center-of-mass inertial system of coordinates.}  Of course, these $3n$ equations for six independent degrees of freedom are too  complicated for practical calculations and analysis. We will find a set of variables more convenient for describing these six degrees of freedom.
First, we single out one vector function with simple dynamics.

\begin{figure}[t] \centering
\includegraphics[width=08cm]{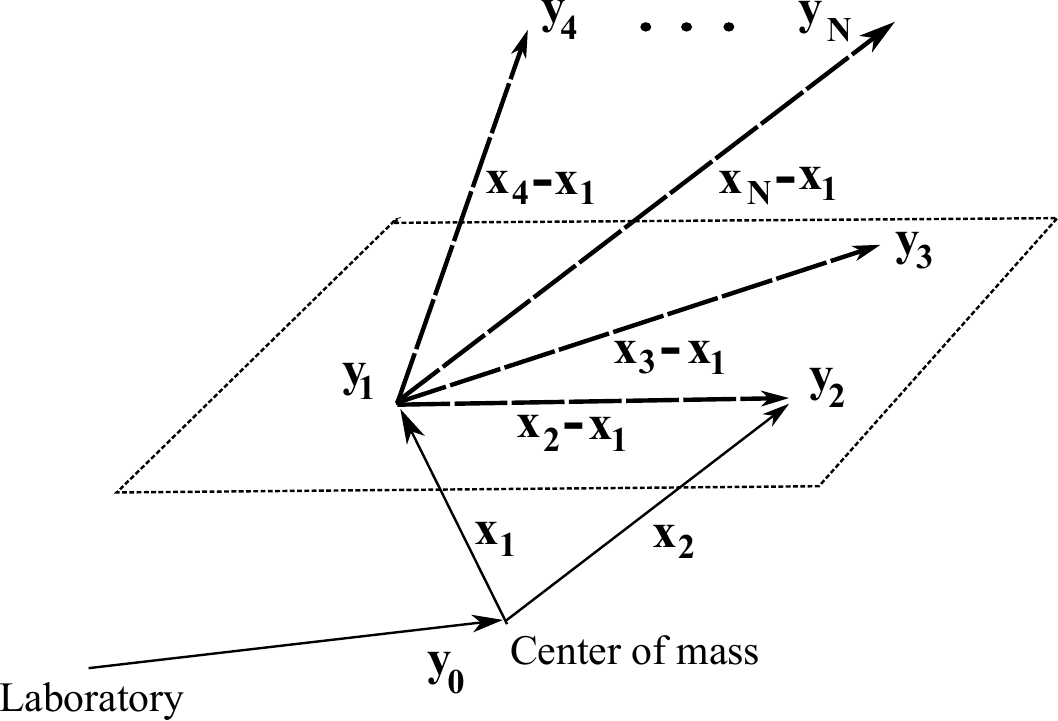}
\caption{Positions of particles with respect to Laboratory, and with respect to the center of mass.}\label{TvT_3}
\end{figure}

Taking the sum of equations (\ref{1.7}), we obtain
\begin{eqnarray}\label{1.8}
\sum_{N=1}^{n}m_N\ddot{\bf y}_N=0.
\end{eqnarray}
So it is convenient to introduce the moving point, called the center of mass of the body, as follows:
\begin{eqnarray}\label{1.9}
{\bf y}_0(t)=\frac1M\sum_{N=1}^{n}m_N{\bf y}_N(t), ~~\mbox{where} ~M=\sum_{N=1}^{n}m_N, ~ \mbox{then} ~ \ddot{\bf y}_0=0.
\end{eqnarray}
Independently of the character of motion of the free body, the center of mass moves along a straight line with constant velocity. Since the initial position and velocity of the body  are assumed to be known, we can compute the initial position and velocity of ${\bf y}_0$, they are ${\bf C}_0=(\sum m_N {\bf c}_N)/M$ and ${\bf V}_0=(\sum m_N {\bf v}_N)/M$. These equalities together with Eq. (\ref{1.9}) determine the center of mass dynamics as follows:   
\begin{eqnarray}\label{1.9.1}
{\bf y}_0(t)={\bf C}_0+{\bf V}_0 t.
\end{eqnarray}
It is convenient to make a change of variables, such that the position vector of the center of mass becomes one of the coordinates of the problem
\begin{eqnarray}\label{1.10}
({\bf y}_1, {\bf y}_2, \ldots , {\bf y}_n ) \rightarrow\left({\bf y}_0=\frac1M\sum_{N=1}^{n}m_N{\bf y}_N(t), ~ {\bf x}_P={\bf y}_P-{\bf y}_0\right), \quad P=1, 2, \ldots , n-1. 
\end{eqnarray}
The variables ${\bf x}_P$ are the position vectors of $n-1$ points of the body with respect to the point of center of mass, see Figure \ref{TvT_3}. The inverse change is
\begin{eqnarray}\label{1.11}
({\bf y}_0,  {\bf x}_1, \ldots , {\bf x}_{n-1}  )\rightarrow \left({\bf y}_P={\bf y}_0+{\bf x}_P,   ~ {\bf y}_n={\bf y}_0-\frac{1}{m_n}\sum_{1}^{n-1}m_N{\bf x}_N\right), \quad P=1, 2, \ldots , n-1. 
\end{eqnarray}

An invertible change of variables can be performed in a Lagrangian, this is known to give an equivalent formulation of the initial problem \cite{Arn_1, Landau_8, deriglazov2010classical}. In terms of new variables (\ref{1.11}), the Lagrangian (\ref{1.6}) reads as follows:
\begin{eqnarray}\label{1.12}
L=\frac12 M\dot{\bf y}_0^2+\frac12\sum_{N=1}^{n}m_N\dot{\bf x}_N^2+\frac12\sum_{A=2}^{4}\sum_{N=2}^{n}\lambda_{AN}\left[({\bf x}_A-{\bf x}_1, {\bf x}_N-{\bf x}_1)-a_{AN}\right],  
\end{eqnarray}
where it was denoted 
\begin{eqnarray}\label{1.13}
{\bf x}_n\equiv-\frac{1}{m_n}\sum_{1}^{n-1}m_N{\bf x}_N,  ~~\mbox{or} ~ \sum_1^n m_N{\bf x}_N=0. 
\end{eqnarray}
This prompts to introduce an independent auxiliary variable ${\bf x}_n$, and to take into account the equality (\ref{1.13}) as one more constraint of the problem, adding it to our action with the corresponding Lagrangian multiplier. According to the results known from classical mechanics, this gives an equivalent Lagrangian 
\begin{eqnarray}\label{1.14}
L=\frac12 M\dot{\bf y}_0^2+\frac12\sum_{N=1}^{n}m_N\dot{\bf x}_N^2+\frac12\sum_{A=2}^{4}\sum_{N=2}^{n}\lambda_{AN}\left[({\bf x}_A-{\bf x}_1, {\bf x}_N-{\bf x}_1)-a_{AN}\right]+\boldsymbol{\lambda}[\sum_1^n m_N{\bf x}_N].   
\end{eqnarray}
It now consist of $3n+3$ independent dynamical variables ${\bf y}_0$ and ${\bf x}_N$, $N=1, 2, \ldots , n$, as well as $3n-3$ auxiliary variables $\lambda_{AN}$ and $\boldsymbol{\lambda}$.  

The center of mass enters only into the first term of the Lagrangian. So, variation of the action with respect to ${\bf y}_0$  gives the equation (\ref{1.9}), whose solution we already know, see  Eq. (\ref{1.9.1}). It is said that three functions $y_0^i$ describe the translational degrees of freedom of a rigid body. Since their evolution is already determined, we omite the first term of (\ref{1.14}) in the subsequent calculations. The remaining variables ${\bf x}_N$ have a simple geometric interpretation. Indeed, let us consider the coordinate system with origin in the center of mass\footnote{According to Eq. (\ref{1.9.1}), it is an inertial system.}, and with axes parallel to the Laboratory axes. This is called the system of center of mass. Then ${\bf x}_N$ defined in Eq. (\ref{1.10}) are just the position vectors of the body's points with respect to this system, see Figure 2. 

Variation of the action (\ref{1.14}) with respect to the variables ${\bf x}$ gives the dynamical equations 
\begin{eqnarray}\label{1.35}
m_1\ddot x_1^i=-\sum_{A,B=2}^4\lambda_{AB}[x_B^i-x_1^i]-\frac12\sum_{A\alpha}\lambda_{A\alpha}[x_A^i+x_\alpha^i-2x_1^i]+m_1\lambda^i, \cr
m_A\ddot x_A^i=\sum_{B=2}^4\lambda_{AB}[x_B^i-x^i_1]+\frac12\sum_{\alpha=5}^n\lambda_{A\alpha}[x_\alpha^i-x_1^i]+m_A\lambda^i, \cr
m_\alpha\ddot x_\alpha^i=\frac12\sum_{A=2}^4\lambda_{A\alpha}[x_A^i-x_1^i]+m_\alpha\lambda^i.
\end{eqnarray}
They are accompanied by the constraints, following from variation of (\ref{1.14}) with respect to $\lambda$ 
\begin{eqnarray}\label{1.36}
\sum_{1}^n m_N{\bf x}_N=0, \qquad ({\bf x}_A-{\bf x}_1, {\bf x}_N-{\bf x}_1)=a_{AN}. 
\end{eqnarray}

\noindent {\bf Integrals of motion.}  Taking scalar product of the equation for $\ddot{\bf x}_N$ of the system (\ref{1.35}) with $\dot{\bf x}_N$, $N=1, 2, \ldots , n$, and then summing all them, 
the right hand side of resulting expression identically vanishes, and we obtain that the rotational energy 
\begin{eqnarray}\label{1.36.1}
E=\frac12 \sum_{N=1}^nm_{N}\dot{\bf x}_N^2,  
\end{eqnarray}
is preserved along any true trajectory of the body: $\frac{dE}{dt}=0$. 

Similarly, taking instead of scalar products the vector products with ${\bf x}_N$, we get that angular momentum of the body with respect to the center of mass 
\begin{eqnarray}\label{4.1}
{\bf m}=\sum_{N=1}^{n}m_N[{\bf x}_N, \dot{\bf x}_N], 
\end{eqnarray}
is preserved\footnote{According to the first Noether's theorem, the integrals of motion could be obtained also using global symmetries of the action (\ref{1.14}). The energy is a consequence of the time-translational invariance, angular momentum is due to rotational invariance, while the constancy of center-of-mass velocity is due to Galileo boosts, see Sect. 7.10 in \cite{deriglazov2010classical}.} as well, $\frac{d{\bf m}}{dt}=0$.

\section{Orthonormal basis rigidly connected to moving body.}\label{SeOB}

Let us show that the distances of the points of the body to the center of mass do not change with time
\begin{eqnarray}\label{1.15}
|{\bf x}_N(t)|=\mbox{const}. 
\end{eqnarray}
This means that the center-of-mass point (that generally is not a point of the body) accompanies the displacement of the body in the space. In turn, Eq. (\ref{1.15}) implies that the angles between vectors ${\bf x}_N(t)$ are preserved. Indeed, consider the vectors ${\bf x}_N(t)$, ${\bf x}_P(t)$ and ${\bf x}_N(t)-{\bf x}_P(t)$ that form a triangle. According to (\ref{1.1}) and (\ref{1.15}), the side lengths of the triangle do not depend on time. Then the same is true for the angles, in particular
\begin{eqnarray}\label{1.16}
({\bf x}_N(t), {\bf x}_P(t))=\mbox{const}. 
\end{eqnarray}
These two properties allow us to imagine the character of possible movements of the body with respect to the center of mass. The movement resembles the evolution of an inclined top. That is, generally,  the body rotates around some axis, one end of which rests at the center of mass, while the other end experiences some movement in space. The exact mathematical formulation of this picture will be given below. 

To show the validity of (\ref{1.15}), we calculate the derivative of $|{\bf x}_N-{\bf x}_P|^2=$\,const, obtaining $({\bf x}_N, \dot{\bf x}_N)+({\bf x}_P,  \dot{\bf x}_P)-(\dot{\bf x}_N, {\bf x}_P)-({\bf x}_N, \dot{\bf x}_P)=0$. Multiplying this expression by $m_P$, summing over $P$ and using Eq. (\ref{1.13}), we get 
\begin{eqnarray}\label{1.17}
M({\bf x}_N, \dot{\bf x}_N)+\sum_{1}^{n}m_P({\bf x}_P,  \dot{\bf x}_P)=0,
\end{eqnarray}
 for any $N$. This implies $({\bf x}_N, \dot{\bf x}_N)=({\bf x}_K,  \dot{\bf x}_K)$, or $({\bf x}_N, \dot{\bf x}_N)=c(t)$, where $c(t)$ is the same for any $N$. Substituting this expression back into (\ref{1.17}), we get $c=0$. So $({\bf x}_N, \dot{\bf x}_N)=\frac12d({\bf x}_N, {\bf x}_N)/dt=0$, 
or $({\bf x}_N, {\bf x}_N)=$\,const, as it was stated.

We now obtain the basic formula, which will allow us to get rid of most of the constraints, and to present the variational problem (\ref{1.14}) in a form convenient for further analysis. Let the basic vectors of the center-of-mass system are the columns ${\bf e}_1=(1, 0, 0)^T$, ${\bf e}_2=(0, 1, 0)^T$ and ${\bf e}_3=(0, 0, 1)^T$. Then
\begin{eqnarray}\label{1.18}
{\bf x}_N(t)={\bf e}_i x_N^i(t), \qquad {\bf x}_N(0)={\bf e}_i x_N^i(0). 
\end{eqnarray}
As $|{\bf x}_N(t)|=$\,const, the vectors ${\bf x}_N(t)$ and ${\bf x}_N(0)$ have the same length, and so are related by some rotation:   $x_N^i(t)=R_{Nij}(t)x_N^j(0)$, where $R^T_N R_N=\bold{1}$ is an othogonal matrix. We will show that this matrix is the same for all particles,  i.e.
\begin{eqnarray}\label{1.19}
x_N^i(t)=R_{ij}(t)x_N^j(0).
\end{eqnarray}
This is the basic formula. It greatly simplifies our task. Indeed, by combining it with the Eqs. (\ref{1.9.1}) and (\ref{1.11}), the evolution of any point of the body can be presented as follows
\begin{eqnarray}\label{1.19.1}
{\bf y}_N(t)={\bf C}_0+{\bf V}_0 t+{\bf x}_N(t)={\bf C}_0+{\bf V}_0 t+R(t){\bf x}_N(0). 
\end{eqnarray}
That is, our task is reduced to finding the equations of motion for three independent dynamical variables contained in the othogonal 
matrix $R(t)$. They are called the rotational degrees of freedom of the rigid body. We emphasize that according to Eq. (\ref{1.19}), initial conditions for the rotation matrix in the theory of a rigid body are fixed once and for all
\begin{eqnarray}\label{1.19.0}
R_{ij}(0)=\delta_{ij}.
\end{eqnarray}
Geometrically, this means that at this instant the columns of the matrix $R$ coincide with basic vectors of center-of-mass system.  The equality (\ref{1.19.1}) is known as the Euler's theorem. 

To prove (\ref{1.19}), we pick  three linearly independent vectors ${\bf x}_A(t)$ among ${\bf x}_N(t)$, and construct the othonormal basis ${\bf R}_i(t)$, rigidly connected with ${\bf x}_A(t)$ at each instant of time. For instance, we can take ${\bf R}_1(t)$ in the direction of ${\bf x}_1(t)$, ${\bf R}_2(t)$ on the plane of the vectors ${\bf x}_1(t)$ and ${\bf x}_2(t)$, and ${\bf R}_3(t)=[{\bf R}_1(t), {\bf R}_2(t)]$. By construction, the vectors ${\bf R}_i$ form an orthonormal basis rigidly connected to the moving body.

As the body fixed basis, we could equally use the vectors 
\begin{eqnarray}\label{1.19.2}
{\bf R}'_j(t)={\bf R}_iU_{ij}(t), 
\end{eqnarray}
where $U$ is time-independent orthogonal matrix, $UU^T=1$. 

Without loss of generality, we can take the basis vectors of laboratory system to coincide with ${\bf R}_i(0)$: ${\bf e}_i={\bf R}_i(0)$. 
Then we can write the following expansions 
\begin{eqnarray}\label{1.20}
{\bf x}_N(t)={\bf R}_i(t)k_N^i, \qquad {\bf x}_N(0)={\bf R}_i(0)k_N^i={\bf e}_ik_N^i. 
\end{eqnarray}
Note that due to Eqs. (\ref{1.15}) and (\ref{1.16}), these  two different vectors have the same coordinates, that were denoted by $k_N^i$. Comparing these expressions with (\ref{1.18}), we conclude that $k_N^i=x_N^i(0)$. Then (\ref{1.20}) and (\ref{1.18}) imply
\begin{eqnarray}\label{1.21}
{\bf e}_jx_N^j(t)={\bf x}_N(t)={\bf R}_i(t)x_N^i(0).
\end{eqnarray}
Two orthonormal basis ${\bf R}_i(t)$ and ${\bf e}_i$ are related by an orthogonal matrix as follows: 
\begin{eqnarray}\label{1.22}
{\bf R}_i(t)={\bf e}_jR_{ji}(t). 
\end{eqnarray}
Substituting this ${\bf R}_i(t)$ into Eq. (\ref{1.21}), we arrive at the desired formula (\ref{1.19}).

Both columns and rows of the matrix $R_{ij}$ have a simple interpretation. Indeed, the last equation states that columns of the matrix $R_{ji}$ coincide with the vectors ${\bf R}_i(t)$ of the body-fixed basis, i.e.
\begin{eqnarray}\label{1.23}
R=({\bf R}_1| {\bf R}_2| {\bf R}_3), \quad \mbox{or} \quad  ({\bf R}_j)_i=R_{ij}.
\end{eqnarray}
Contracting  Eq. (\ref{1.19}) with ${\bf e}_i$, this can be presented in vector form 
\begin{eqnarray}\label{1.23.1}
{\bf x}_N(t)={\bf R}_j(t)x_N^j(0). 
\end{eqnarray}
This has a simple meaning: points of the body are at rest with respect to the basis ${\bf R}_j(t)$. 

The interpretation of the rows of the matrix $R_{ij}$ becomes clear if we invert Eq. (\ref{1.22}) as 
follows: ${\bf e}_i=R_{ij}(t){\bf R}_j(t)$.  So the rows 
\begin{eqnarray}\label{1.23.1.1}
R=\left(
\begin{array}{c}
{\bf G}_1\\
{\bf G}_2 \\
{\bf G}_3
\end{array}
\right), 
\end{eqnarray}
represent the laboratory basis vectors ${\bf e}_i$ in the rigid body basis. For instance, the numbers ${\bf G}_1(t)=(R_{11}, R_{12}, R_{13})$ are components of the basis vector ${\bf e}_1$ in the basis ${\bf R}_i(t)$.   In terms of the rows ${\bf G}_i(t)$, Eq. (\ref{1.19}) reads
\begin{eqnarray}\label{1.23.2}
x_N^i(t)=({\bf G}_i(t), {\bf x}_N(0)), 
\end{eqnarray}
that is the coordinates $x_N^i(t)$ of a point of the body are projections of the initial vector of position on the vectors ${\bf G}_i(t)$.

\section{Angular velocity, mass matrix and tensor of inertia.}\label{AVM} 
We will now obtain {\it kinematic} consequences of the formula (\ref{1.19}), and introduce some quantities that will be convenient for describing a body in the center-of-mass system: various forms of angular velocity, mass matrix and tensor of inertia. We point out that all them automatically arise also as the phase space quantities when analyzing the Hamiltonian equations of motion, see Sect. \ref{SeHE}. 

\begin{figure}[t] \centering
\includegraphics[width=06cm]{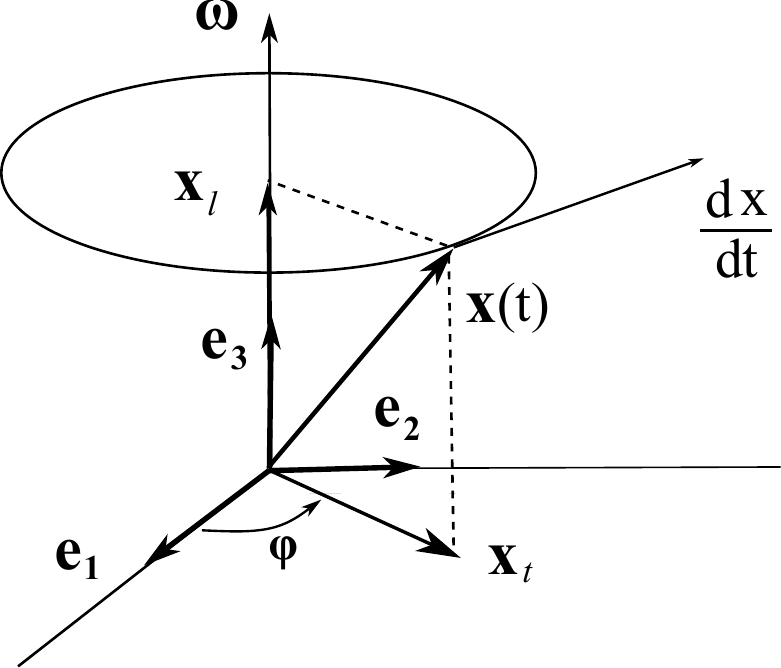}
\caption{Instantaneous angular velocity $|\boldsymbol{\omega}|=\frac{d\varphi}{dt}$ of precession.}\label{TvT_4}
\end{figure}

{\bf Instantaneous angular velocity of a rigid body.} 
Derivative of Eq. (\ref{1.19}) can be presented in various forms as follows:
\begin{eqnarray}\label{1.24}
\dot x_N^i(t)=\dot R_{ij}x_N^j(0)=-\hat\omega_{ij}(t)x_N^j(t)=\epsilon_{ikj}\omega_k(t)x_N^j(t).
\end{eqnarray}
Here 
\begin{eqnarray}\label{1.25}
\hat\omega_{ij}(t)=-(\dot RR^T)_{ij},
\end{eqnarray}
is an antisymmetric matrix, while $\omega_k$ is the corresponding vector (see Eq. (\ref{0.2}))
\begin{eqnarray}\label{1.26}
\omega_k(t)\equiv\frac12\epsilon_{kij}\hat\omega_{ij}=-\frac12\epsilon_{kij}(\dot RR^T)_{ij}, \qquad 
\hat\omega_{ij}=\epsilon_{ijk}\omega_k.
\end{eqnarray}
It is called instantaneous angular velocity of the body. Eq. (\ref{1.24}) in the vector form reads
\begin{eqnarray}\label{1.27}
\dot{\bf x}_N=[\boldsymbol{\omega}, {\bf x}_N].
\end{eqnarray}
This implies that velocity of any point $N$ is orthogonal to the plane of the vectors $\boldsymbol{\omega}$ and ${\bf x}_N$. Besides we have 
$|{\bf x}_N|=\mbox{const}$, as it should be according to (\ref{1.15}). 
When $\boldsymbol{\omega}$ does not depend on time, this equation describes precession of the vector ${\bf x}_N$ around the 
axis $\boldsymbol{\omega}$, see Figure \ref{TvT_4}. Indeed, let us place the beginning of the vector $\boldsymbol{\omega}$ at the origin of center-of-mass system. Let ${\bf x}_N(0)={\bf x}_{N\Vert}(0)+{\bf x}_{N\bot}(0)$ is decomposition of initial position on longitudinal and transverse parts with respect to $\boldsymbol{\omega}$, see Figure \ref{TvT_4}. Then 
\begin{eqnarray}\label{1.28}
{\bf x}_N(t)={\bf x}_{N\Vert}(0)+{\bf x}_{N\bot}(t)={\bf x}_{N\Vert}(0)+|{\bf x}_{N\bot}(0)|\left[{\bf e}_1\cos(|\boldsymbol{\omega}|t)+
{\bf e}_2\sin(|\boldsymbol{\omega}|t)\right]
\end{eqnarray}
is a solution to Eq. (\ref{1.27}).  The point ${\bf x}_N$ describe  a circle around $\boldsymbol{\omega}$ with the frequency of rotation (or angular velocity\footnote{Let $\varphi(t)$ is the angle of rotation, see Figure \ref{TvT_4}. The angular velocity is related with the linear velocity as follows: $\dot\varphi=|\dot{\bf x}_{N\bot}|/|{\bf x}_{N\bot}|$. Using (\ref{1.28}), we get $\dot\varphi=|\boldsymbol{\omega}|$.}) equal  to magnitude of this vector $|\boldsymbol{\omega}|$.   When $\boldsymbol{\omega}$ is a function of time, the end of this vector experiences some movement, and the described precession is only  a part of the total moviment. 

The basic vectors ${\bf R}_j$, being rigidly connected with the body, precess according the same rule\footnote{Note that substituting (\ref{1.22}) and (\ref{1.26}) into (\ref{1.27.1}), we get the identity $\dot R_{ij}=\dot R_{ij}$.}
\begin{eqnarray}\label{1.27.1}
\dot{\bf R}_j=[\boldsymbol{\omega}, {\bf R}_j].
\end{eqnarray}
Note that $\hat\omega=\dot RR^T$ and $R^T\dot R$ are two different antisymmetric matrices. So we denote 
\begin{eqnarray}\label{1.29}
\hat\Omega_{ij}=-(R^T\dot R)_{ij}, \qquad \Omega_k\equiv\frac12\epsilon_{kij}\hat\Omega_{ij}=-\frac12\epsilon_{kij}(R^T\dot R)_{ij}, \qquad \mbox{then} \quad \hat\Omega_{ij}=\epsilon_{ijk}\Omega_k.  
\end{eqnarray}
These definitions imply the relations
\begin{eqnarray}\label{1.30}
\hat\Omega=-R^T \hat\omega R, \qquad \omega_i=R_{ij}\Omega_j. 
\end{eqnarray}
The functions $\Omega_j(t)$ are called components of {\it angular velocity in the body} \cite{Arn_1}. Their meaning is clear from the following line: 
\begin{eqnarray}\label{1.30.0}
{\boldsymbol\omega}\equiv(\omega_1, \omega_2, \omega_3)^T=
{\bf e}_i\omega_i={\bf e}_iR_{ij}\Omega_j={\bf R}_j(t)\Omega_j.
\end{eqnarray}
That is the numbers $\Omega_j$ are coordinates of the angular velocity vector $\boldsymbol\omega={\bf e}_i\omega_i$ with respect to the orthonormal basis ${\bf R}_j(t)$, rigidly connected to the body. We emphasise that there is no the independent vector ${\boldsymbol\Omega}$ in the formalism. Sometimes we will write $I{\boldsymbol\Omega}$ to denote the quantities $I_{ij}\Omega_j$ and so on, which is certain abuse of notation.

{\bf Dynamics of a body in the case of constant angular velocity.} When the angular velocity ${\boldsymbol\omega}$ is known to be time independent, we can combine the solution (\ref{1.28}) to the equation $\dot{\bf x}_N=[\boldsymbol{\omega}, {\bf x}_N]$  with (\ref{1.19.1}) and get 
\begin{eqnarray}\label{1.30.1}
{\bf y}_N(t)={\bf C}_0+{\bf V}_0 t+{\bf x}_{N\Vert}(0)+|{\bf x}_{N\bot}(0)|\left[{\bf e}_1\cos(|\boldsymbol{\omega}|t)+
{\bf e}_2\sin(|\boldsymbol{\omega}|t)\right].  
\end{eqnarray}
That is in  this case the problem of the motion of a rigid body can be considered already solved. 

Consider some point of the body, that at $t_0$ is located on the axis $\boldsymbol{\omega}$. Then Eq. (\ref{1.30.1}) implies, that this point will be located on the axis at all future instants of time. That is when $\dot{\boldsymbol{\omega}}=0$, this axis can be considered as rigidly connected with the body.  If the axis ${\boldsymbol\omega}$ moves in space, it must also move in the body, see the end of Sect. \ref{FOE}.

In the general case of time-dependent angular velocity, the equation $\dot{\bf x}_N=[\boldsymbol{\omega}, {\bf x}_N]$  turns out to be much less useful. The specific property of the theory is
that equations on $\Omega_i$ turn out to be closed, in the sense that they involve only $\Omega_i$ itself. These are the famous Euler equations, see below. From (\ref{1.30}) it follows, that to determine ${\boldsymbol\omega}$ we need also to know $R(t)$. In turn, equations for $R(t)$ 
involve $\Omega_i$ but do not involve ${\boldsymbol\omega}$. Therefore, it turn out to be more simple task to solve the system of equations for $R_{ij}$ 
and $\Omega_i$, which do not involve ${\boldsymbol\omega}$ at all, and use the obtained $R(t)$ to find the motion by the 
formula ${\bf x}_N(t)=R(t){\bf  x}_N(0)$, instead of using (\ref{1.27}). On other hand, the angular velocity has simple geometric meaning as the axis of instaneous rotation, and will be useful for visualization of the free rigid body motion, see Sect. \ref{SePo}.

In resume, when angular velocity is a constant vector, it turns out to be the basic variable for determining the motion.  In general case, the variables $\Omega_i$ and $R_{ij}$ are more convenient, as equations of motion of the rigid body are formulated in terms of these variables.

{\bf Mass matrix, tensor of inertia, and their properties under rotations of center-of-mass system.} Second term of the Lagrangian (\ref{1.14}) represents the kinetic energy of body's rotation\footnote{More exactly, it is kinetic energy of the body with respect to the center of mass.}.  As we have shown above, the energy preserve its value along solutions of equatios of motion. 
Using Eqs.  (\ref{4.1}), (\ref{1.19}) and (\ref{1.27}), the energy can be presented in various forms as follows: 
\begin{eqnarray}\label{1.31_1}
E=\frac12\sum_{N=1}^{n}m_N\dot{\bf x}_N^2=
\frac12g_{ij}\dot{\bf R}_{i}\dot{\bf R}_{j}
=\frac12I_{ij}\Omega_i\Omega_j,
\end{eqnarray}
In Eq. (\ref{1.31_1}) appeared two numeric matrices. The symmetric non degenerate matrix $g$ with the components 
\begin{eqnarray}\label{1.32}
g_{ij}\equiv\sum_{N=1}^{n}m_Nx_N^i(0)x_N^j(0),
\end{eqnarray}
will be called mass matrix,  while the symmetric matrix $I$ with the components 
\begin{eqnarray}\label{1.33}
I_{ij}\equiv\sum_{N=1}^{n}m_N\left[{\bf x}_N^2(0)\delta^{ij}-x_N^i(0)x_N^j(0)\right]=\left[g_{kk}\delta_{ij}-g_{ij}\right],
\end{eqnarray}
is called tensor of inertia. In the limit of continuous distribution of particles with mass density $\rho({\bf x})$, the sum in these expressions should be replaced by integral, for instance
\begin{eqnarray}\label{1.34}
g_{ij}=\int d^3x\rho({\bf x})x^ix^j.
\end{eqnarray}
These two time independent matrices are characteristics of spatial distribution of masses in the body at the initial instance of time. They are not invariant under translations: they were defined,  and should be computed in the center-of-mass system. 

Besides, the explicit form of these numeric matrices depends on the initial position of the body. Equivalently, it can be said that they change when we pass from one Laboratory basis to another one, related by some rotation. Mathematically, they transform as tensors under rotations of the center-of-mass system. Indeed, consider two orthonormal bases related by rotation with help of numeric orthogonal matrix $U^TU=1$: ${\bf e}'_i={\bf e}_kU^T_{ki}$. Coordinates of the body's particles in these bases are related as follows: 
$x'^i=U_{ij}x^j$.
Then Eq. (\ref{1.32}) implies that the matrices $g'_{ij}$ and $g_{ij}$, computed in these bases, are related by 
\begin{eqnarray}\label{1.33.2}
g'_{ij}\equiv\sum_N m_Nx'^i_N(0)x'^j_N(0)=U_{ia}(\sum_N m_Nx_N^a(0)x_N^b(0))U_{jb}=U_{ia}g_{ab}U_{jb}, \quad \mbox{or} \quad g'=UgU^T.
\end{eqnarray}
The inertia tensor has the same transformation rule. 

Let us prove that $g$ is non degenerate matrix. According to the linear algebra, given a symmetric matrix $g$, there is an orthogonal matrix $U$ such 
that $UgU^T=\tilde g=\mbox{diagonal}(g_1, g_2, g_3)$, or $\sum m_N(U{\bf x}_N)^i(U{\bf x}_N)^j=\mbox{diagonal}(g_1, g_2, g_3)$, then 
\begin{eqnarray}\label{1.34.1}
\det g=\det\tilde g=g_1g_2g_3, \qquad g_i=\sum_N m_N(U{\bf x}_N)^i(U{\bf x}_N)^i, \qquad \mbox{without summation over $i$}. 
\end{eqnarray}
Since our body has four particles not lying on the same plane, among ${\bf x}_N$ there are three linearly independent vectors. Together with Eq. (\ref{1.34.1}) this implies $g_i>0$ and $\det g>0$.  

If one of $g_i$, say $g_1=0$, this implies $x_N^1=0$ for any $N$, that is we have a plane body. Similarly, $g_1=g_2=0$ implies that the body is a solid rod. 

Recall that a symmetric non degenerate matrix $g$ has three orthogonal eigenvectors with non vanishing eigenvalues: $g{\bf b}_i=\lambda_i{\bf b}_i$. They can be chosen to be of unit length, and forming a right-handed triple. Applying the defined above matrix $U$ to this equality, we get $\tilde g(U{\bf b})_i=\lambda_i(U{\bf b})_i$, which implies that the eigenvalues coincide with diagonal elements of $\tilde g$, $\lambda_i=g_i$. 
According to Eq. (\ref{1.33}), ${\bf b}_i$ are also eigenvectors of $I$, with the eigenvalues, say, $I_1, I_2, I_3$. The straight lines determined by the vectors ${\bf b}_i$ are called principal axis of inertia of the body, while the numbers $I_j$ are principal moments of inertia. By construction, the axes are rigidly connected with the body. 

For the latter use, we present the relations among the eigenvalues
\begin{eqnarray}\label{1.35.1}
2g_1=I_2+I_3-I_1, \quad 2g_2=I_1+I_3-I_2, \quad 2g_3=I_1+I_2-I_3, \cr
I_1=g_2+g_3, \quad I_2=g_1+g_3, \quad I_3=g_1+g_2. \qquad \qquad 
\end{eqnarray}
Note that $g_i>0$ implies a number of consequences: (a) $I_i>0$; (b) $g_1=g_2=g_3$ implies $I_1=I_2=I_3=2g_1$; (c) the sum of any two moments of inertia is always not less than the third, for instance $I_2+I_3\ge I_1$.  For a plane body, say $g_1=0$, we get $I_1=I_2+I_3=g_2+g_3$. For a solid rod, say $g_1=g_2=0$, we get $I_3=0$, $I_1=I_2=g_3$. 

\label{TPGI}
Without loss of generality, we can assume that the matrices $g$ and $I$ in (\ref{1.31_1}) are diagonal matrices. Indeed, 
let $g$ in Eq. (\ref{1.31_1}) is not diagonal and let $U$ is its diagonalizing matrix, $UgU^T=\tilde g=\mbox{diagonal}(g_1, g_2, g_3)$. Let's turn the laboratory basis with help of $U^T$. According to Eq. (\ref{1.33.2}), calculating the kinetic energy in this basis, we arrive at Eq. (\ref{1.31_1}) with  diagonal matrices $g$ and $I$. Geometrically this means that at $t=0$ we have chosen the Laboratory axes to coincide with the 
principal exes of inertia of the body. Besides, from equations (\ref{1.19}) and (\ref{1.19.0}) it follows, that the body fixed frame ${\bf R}_j$ at $t=0$ also coincides with these two bases. Since the inertia axes and the body frame axes are rigidly connected with the body, they will coincide with each other at all future instants of time: ${\bf R}_j(t)={\bf b}_j(t)$.  

Below, we work with an asymmetric rigid body, that is $I_1\ne I_2\ne I_3$, assuming that that the matrices $g$ and $I$ are diagonal. This implies that at $t=0$ the Laboratory axes has been fixed in the directions of inertia axes of the body. This has an important consequence: we have no more a freedom to rotate the Laboratory system with the aim to simplify the equations of motion. The case of a symmetric body $I_1= I_2\ne I_3$ will be discussed in Sect. \ref{IN}.   

{\bf Angular momentum and angular momentum in the body.}  It is convenient to introduce the components $M_k$ of the angular momentum ${\bf m}$ in the body-fixed basis ${\bf R}_k$ as follows: 
\begin{eqnarray}\label{4.6}
{\bf m}=(m_1, m_2, m_3)^T={\bf e}_i m_i={\bf e}_j R_{jk}R^T_{ki}m_i={\bf R}_kR^T_{ki}m_i\equiv{\bf R}_kM_k, 
\end{eqnarray}
that is $M_k=R^T_{ki}m_i$. $M_k$ are called {\it components of angular momentum in the body} \cite{Arn_1}.  

Using the basic formula (\ref{1.19}) in the definition of angular momentum (\ref{4.1}), it can be presented  in various forms as follows: 
\begin{eqnarray}\label{2.8}\label{s2.4}
{\bf m}=\sum_{N=1}^{n}m_N[{\bf x}_N, \dot{\bf x}_N]=\sum_i g_i[{\bf R}_i, \dot{\bf R}_i]=RIR^T{\boldsymbol\omega}=RI{\bf\Omega}=R{\bf M}.   
\end{eqnarray}
In obtaining the third and fourth equalities we used Eqs. (\ref{0.2}), (\ref{1.29}), (\ref{1.33}) and (\ref{1.30}). 

In the Laboratory basis, the vector ${\bf m}$ is constant, while the end of the vector ${\boldsymbol\omega}$ moves along a complex trajectory. The  formula $M_k=I_{kj}\Omega_j$ shows that  coordinates of ${\bf m}$ and ${\boldsymbol\omega}$ in the body-fixed frame are rigidly conected.

Similarly, the kinetic energy (\ref{1.31_1}) can be presented in various forms  as follows: 
\begin{eqnarray}\label{1.31}
E=\frac12\sum_{N=1}^{n}m_N\dot{\bf x}_N^2=
\frac12g_{ij}\dot{\bf R}_{i}\dot{\bf R}_{j}=\frac12 ({\boldsymbol\omega}, {\bf m})
=\frac12 (RIR^T)_{ij}\omega_i\omega_j
=\frac12I_{ij}\Omega_i\Omega_j=\frac12(RI^{-1}R^T)_{ij}m_im_j=\frac12 I^{-1}_{ij}M_i M_j,
\end{eqnarray}
Note that for any motion we have $({\boldsymbol\omega}, {\bf m})>0$, that is the vectors of conserved angular momentum and of angular velocity always form an acute angle. 

We emphasize that Eqs. (\ref{2.8}) and (\ref{1.31})  are purely kinematic in nature. Therefore, they are also applicable in the cases of a body with a fixed point, as well as a body under the action of external forces.

\section{Action functional and second-order Lagrangian equations for rotational degrees of freedom.}\label{SeEL}
Let us return to the discussion of equations of motion (\ref{1.35}) and (\ref{1.36}), implied by the Lagrangian (\ref{1.14}). 
As we saw above, any solution of this system is of the form $x_N^i(t)=R_{ij}(t)x_N^j(0)$. Substituting this anzatz into the constraints (\ref{1.36}), we  just learn that they should be satisfied at the initial instant of time. Substituting the anzatz into the dynamical equations, we obtain $3n$ equations that contain $R_{ij}$ and its second derivatives. Multipluing the equation with number $N$ by $x_N^j(0)$  and taking their sum, we obtain the following equations for determining $R_{ik}$: 
\begin{eqnarray}\label{1.37}
\ddot R_{ik}g_{kj}=-R_{ik}\lambda_{kj}, \quad \mbox{they are accompanied by the constraints} ~ R^TR={\bf 1}.
\end{eqnarray}
By $\lambda_{jk}(t)$ in Eq. (\ref{1.37})  was denoted the following symmetric matrix
\begin{eqnarray}\label{1.38}
\lambda_{jk}=-\sum_{AB}\lambda_{AB}\left[x_1^jx_1^k+x_A^jx_B^k-x_B^{(j}x_1^{k)}\right]-\frac12\sum_{A\alpha}\lambda_{A\alpha}\left[x_\alpha^{(j}x_A^{k)}-x_\alpha^{(j}x_1^{k)}-x_A^{(j}x_1^{k)}-2x_1^{j}x_1^{k}\right],
\end{eqnarray}
where all $x_N^i$ are taken at the instant $t=0$. Note that it depends on the unknown dynamical variables $\lambda_{AN}(t)$.
A remarcable property of the system (\ref{1.37}) is that we do not need to know  $\lambda_{jk}(\lambda_{AN}(t))$ to solve it. As we show below, $\lambda_{jk}$ are uniquely determined by the system (\ref{1.37}) itself. It determines $\lambda_{jk}$ algebraically, as some functions of $R$ and $\dot R$. Substitution of these functions back into the Eq. (\ref{1.37}) gives a well posed Cauchy problem for determining $R(t)$. Let us see how all this works.

{\bf Variational problem for the equations (\ref{1.37}).}
Here we show that  the system (\ref{1.37}) follows from the variational problem, in which $\lambda_{jk}$ are just the Lagrangian multipliers for the constraints $R^TR={\bf 1}$.

Consider a dynamical system with configuration space variables $R_{ij}(t)$, $\lambda_{ij}(t)$, $i, j=1, 2, 3$, where $R$ is $3\times 3$ matrix 
and $\lambda$ is symmetric $3\times 3$ matrix. Let $g_{ij}=\mbox{diag}(g_1, g_2, g_3)$ is diagonal numeric matrix. Then the Lagrangian action
\begin{eqnarray}\label{2.1}
S=\int dt ~ ~ \frac12 g_{ij}\dot R_{ki}\dot R_{kj} -\frac12 \lambda_{ij}\left[R_{ki}R_{kj}-\delta_{ij}\right]\equiv\int dt ~~\frac12 \mbox{tr}[{\dot Rg\dot R^T]-
\frac12\mbox{tr}[\lambda(R^TR-{\bf 1})}], 
\end{eqnarray}
implies both dynamical equations and costraints (\ref{1.37}) as the conditions of extremum of this variational problem. In particular, variation of the action with respect to $\lambda$ implies the constraints $R^TR={\bf 1}$. They mean that $R(t)$ is an element of the group of rotations $SO(3)$. It is said, that the variational problem is formulated for a point moving on the group manifold $SO(3)$. 

It should be noted that when formulating a variational problem in classical mechanics, we usually look for the extremum of the functional $\int dt L(q, \dot q)$ for arbitrarily chosen initial and final positions: $q(0)=q_0$, $q(t_1)=q_1$. In the case of a rigid body, the initial position of the problem is fixed once and for all according to the equation (\ref{1.19.0}): $R_{ij}(0)=\delta_{ij}$.

The problem (\ref{2.1}) has also a simple mechanical interpretation.  We can rewrite (\ref{2.1}) in terms of columns ${\bf R}_j$ of the matrix $R_{ij}$ as follows: 
\begin{eqnarray}\label{2.2}
L=\frac12\left[g_1\dot{\bf R}_1^2+g_2\dot{\bf R}_2^2+g_3\dot{\bf R}_3^2\right]-\frac12 \lambda_{ij}\left[({\bf R}_{i}, {\bf R}_{j})-\delta_{ij}\right]. 
\end{eqnarray}
As was proved above, $g_i>0$. So the variational problem describes three particles of masses $g_i$, which are connected by massless solid rods of the length equal to $\sqrt 2$, and move freely on the surface of sphere with unit radius.

{\bf Second-order equations of motion for $R_{ij}$.} Variation of the action (\ref{2.2}) with respect to ${\bf R}_i$ and $\lambda_{ij}$ gives the equations of motion (there is no summation over $i$ in Eq. (\ref{2.3}) ) 
\begin{eqnarray}\label{2.3}
g_i\ddot{\bf R}_i=-\sum_j\lambda_{ij}{\bf R}_j, 
\end{eqnarray}
\begin{eqnarray}\label{2.4}
({\bf R}_i, {\bf R}_j)=\delta_{ij}.
\end{eqnarray}
The auxiliary variables $\lambda_{ij}$ can be excluded from the second-order equations (\ref{2.3}) as follows. Calculating first and second derivatives of the constraint (\ref{2.4}) we get the consequences  
\begin{eqnarray}\label{2.3.1}
(\dot{\bf R}_i, {\bf R}_j)+({\bf R}_i, \dot{\bf R}_j)=0, \qquad (\ddot{\bf R}_i, {\bf R}_j)+({\bf R}_i, \ddot{\bf R}_j)+2(\dot{\bf R}_i, \dot{\bf R}_j)=0. 
\end{eqnarray}
Using Eqs. (\ref{2.3}) for second derivatives in the last expression, we get  
\begin{eqnarray}\label{2.5}
\lambda_{ij}=\frac{2g_i g_j}{g_i+g_j}(\dot{\bf R}_i, \dot{\bf R}_j). 
\end{eqnarray}
Using them in (\ref{2.3}), we obtain closed system of second-order equations for determining the temporal evolution of rotational degrees of freedom of the body
\begin{eqnarray}\label{2.6}
\ddot{\bf R}_i=-\sum_j\frac{2g_j}{g_i+g_j}(\dot{\bf R}_i, \dot{\bf R}_j){\bf R}_j, \qquad  ({\bf R}_i, {\bf R}_j)=\delta_{ij}. 
\end{eqnarray}
They should be solved with initial conditions $R_{ij}(0)=\delta_{ij}$, $\dot R_{ij}(0)=V_{ij}$, $V_{ij}=-V_{ji}$. They follow from Eqs. (\ref{1.22}) and (\ref{2.3.1}). 

We emphasize once again that not all solutions to the equations (\ref{2.6}) with the diagonal mass matrix $g$  describe the possible motions of a rigid body. Let $R_{ij}(t)$ be a solution of (\ref{2.6}). According to (\ref{1.19}), by construction of the variables $R_{ij}(t)$, this describes the possible motion of the rigid body only if at some instant of time, say $t=0$, the solution passes through the unit element of $SO(3)$: $R_{ij}(0)=\delta_{ij}$. Then this $R_{ij}(t)$ corresponds to the motion of our rigid body, which at the moment $t=0$ had axes of inertia in the direction of the axes of the laboratory.

{\bf The rotational energy is not an independent integral of motion.} The equations of motion (\ref{2.6}) imply conservation of energy and angular momentum. Taking scalar product of (\ref{2.6}) with the vector $g_i\dot{\bf R}_i $ and summing over $i$ we get the conservation 
of energy  
\begin{eqnarray}\label{2.7}
\frac{dE}{dt}=0, \quad \mbox{where} \quad E=\frac12\sum_i g_i\dot{\bf R}_i^2. 
\end{eqnarray}
The energy can be presented in various forms, see (\ref{1.31}). 
Similarly, using the vector product instead of the scalar product, we get the conservation of angular momentum
\begin{eqnarray}\label{s2.4_1}
\frac{d{\bf m}}{dt}=0, \quad \mbox{where} \quad {\bf m}=\sum_i g_i[{\bf R}_i, \dot{\bf R}_i].
\end{eqnarray}
As it should be, this conserved vector coincides with that defined in (\ref{s2.4}) 
\begin{eqnarray}\label{4.3}
m_i=\sum_{N=1}^{n}m_N[{\bf x}_N, \dot{\bf x}_N]_i=\sum_{N=1}^{n}m_N[{\bf x}_N, [{\boldsymbol\omega}, {\bf x}_N]]_i=
\sum_{N=1}^{n}m_N(\omega^i({\bf x}_N,  {\bf x}_N)-x_N^i({\boldsymbol\omega},  {\bf x}_N))= \qquad \qquad   \cr \sum_{N=1}^{n}m_N[({\bf x}_N(0), {\bf x}_N(0))\delta_{ij}- R_{ia}R_{jb}x_N^a(0) x_N^b(0)]\omega_j=
\sum_{N=1}^{n}m_N[{\bf x}_N^2(0)\delta_{ab}-x_N^a(0) x_N^b(0)]R_{ia}(R^T{\boldsymbol\omega})_b= \cr  R_{ia}I_{ab}(R^T{\boldsymbol\omega})_b.  \qquad \qquad \qquad \qquad \qquad \qquad \qquad \qquad \qquad \qquad  
\end{eqnarray}
Here we used Eqs.  (\ref{1.27}), (\ref{1.19}), (\ref{1.33}) and (\ref{1.30}). 

Eqs. (\ref{2.8}) and (\ref{1.19.0}) imply an important consequence: the initial dates for the angular velocity cannot be taken arbitrary, but are fixed by the conserved angular momentum
\begin{eqnarray}\label{4.3.0}
m_i=I_{ij}\omega_j(0)=I_{ij}\Omega_j(0).  
\end{eqnarray}
Then the expression for the energy $E=\frac12\sum_i I_i\Omega_i^2(t)=\frac12\sum_i I_i\Omega_i^2(0)$ implies\footnote{In the covariant form this is $E=\frac12 I^{-1}_{ij}m_im_j$, see also (\ref{1.31}).}
\begin{eqnarray}\label{2.8.1}
E=\frac12\sum_i \frac{1}{I_i} m_i^2,
\end{eqnarray}
that is the rotational energy of a free rigid body does not represent an independent integral of motion\footnote{This should be compared with Sect. 28 of \cite{Arn_1}.}.

\section{\bf First-order form of equations of motion and the Euler-Poisson equations.}\label{FOE} 
The vector equation (\ref{2.6}) of second order is equivalent to a system of two equations of first order for twice the number of independent variables. 
To obtain the system, consider the space of {\it mutually independent} dynamical variables $R_{ij}(t)$ and $\Omega_i(t)$, subject to the equations $({\bf R}_i, {\bf R}_j)=\delta_{ij}$ as well as to
\begin{eqnarray}
\ddot R_{aj}=-\sum_k\frac{2g_k}{g_j+g_k}(\dot{\bf R}_j, \dot{\bf R}_k)R_{ak}, \label{2.9.1} \\  
\Omega_k=-\frac12\epsilon_{kij}(R^T\dot R)_{ij}. \label{2.9.2}
\end{eqnarray}
That is $R_{aj}(t)$ satisfies the equations (\ref{2.6}), while $\Omega_k(t)$ accompanies the dynamics of $R_{aj}(t)$ according to (\ref{2.9.2}).  Evidently, this system is equivalent to (\ref{2.6}). Multiplying Eq. (\ref{2.9.1}) on the invertible matrix $R_{ai}$ we get
\begin{eqnarray}\label{2.9}
({\bf R}_i, \ddot{\bf R}_j)=-\frac{2g_i}{g_i+g_j}(\dot{\bf R}_i, \dot{\bf R}_j). 
\end{eqnarray}
Let us separate symmetric and antisymmetric parts of Eq. (\ref{2.9}) as follows:
\begin{eqnarray}
({\bf R}_i, \ddot{\bf R}_{j})+({\bf R}_j, \ddot{\bf R}_{i}) =-2(\dot{\bf R}_i, \dot{\bf R}_j), \label{2.9.3} \\ 
({\bf R}_i, \ddot{\bf R}_{j})-({\bf R}_j, \ddot{\bf R}_{i})=-2\frac{g_i-g_j}{g_i+g_j}(\dot{\bf R}_i, \dot{\bf R}_j). \label{2.9.4} 
\end{eqnarray}
According to (\ref{2.3.1}), the equation (\ref{2.9.3})  is a consequence of $({\bf R}_i, {\bf R}_j)=\delta_{ij}$, and can be omitted from the system. 
Further, the scalar product on r.h.s. of Eq. (\ref{2.9.4}) can be identically rewritten in terms of $\Omega_i$ (\ref{2.9.2}) as follows\footnote{Note that the matrix $N$ has the properties $N_{ij}\Omega_j=0$, $N_{ik}N_{kj}=N_{ij}$, and when acting on an arbitrary vector, $N$ projects it onto a plane orthogonal to $\Omega_i$. Acting on three vectors according to the rule: $N_{ij}{\bf B}_j\equiv {\bf C}_i$, we obtain three coplanar vectors: $\Omega_i{\bf C}_i=0$.}:
\begin{eqnarray}\label{2.10}
(\dot{\bf R}_i, \dot{\bf R}_j)=\Omega^2N_{ij}(\Omega), \qquad \mbox{where} \quad N_{ij}(\Omega)=
\delta_{ij}-\frac{\Omega_i \Omega_j}{\Omega^2}, \quad \Omega^2\equiv\sum_i\Omega_i^2.  
\end{eqnarray}
Due to the identification (\ref{0.2}), we can contract the antisymmetric equation (\ref{2.9.4}) with $-\frac12\epsilon_{kij}$, obtaining an equivalent equation. Using (\ref{2.9.2}) and (\ref{2.10}), this can be presented as the first-order equations for determining $\Omega_i$
\begin{eqnarray}\label{2.11}
\dot\Omega_k=-\sum_{ij}\epsilon_{kij}\frac{g_i}{g_i+g_j}\Omega_i\Omega_j. 
\end{eqnarray}
For the components they read 
\begin{eqnarray}\label{2.12}
\dot\Omega_1=\frac{1}{I_1}(I_2-I_3)\Omega_2\Omega_3, \cr
\dot\Omega_2=\frac{1}{I_2}(I_3-I_1)\Omega_1\Omega_3, \cr
\dot\Omega_3=\frac{1}{I_3}(I_1-I_2)\Omega_1\Omega_2.
\end{eqnarray}
These are the famous Euler equations. 
Here $I_i$ are components of the tensor of inertia, see (\ref{1.35.1}).  In a more compact form, with use of vector product they read
\begin{eqnarray}\label{2.13}
I\dot{\boldsymbol\Omega}=[I{\boldsymbol\Omega},{\boldsymbol\Omega}].  
\end{eqnarray}
The equation (\ref{2.9.2}) can  be rewritten in the form of first-order equation for determining of $R_{ij}$
\begin{eqnarray}\label{2.14} 
\dot R_{ij}=-\epsilon_{jkm}\Omega_k R_{im}. 
\end{eqnarray}
Using the dual basis ${\bf G}_i$, they also can be written in the vector form as follows: $\dot{\bf G}_i=-[{\boldsymbol\Omega}, {\bf G}_i]$. That is the dual basis precess around the vector of angular velocity in the body. 
For the case of heavy top,  three of these equations (namely the equations for the matrix elements $R_{31}, R_{32}$ and $R_{33}$, 
that is $\dot{\bf G}_3=[{\bf G}_3, {\boldsymbol\Omega}]$) were obtained by Poisson \cite{Poi_1842}, 
and bear his name, see the historical notes in \cite{Lei_1965}. So we will call the equations (\ref{2.13}) and (\ref{2.14}) the Euler-Poisson equations. 

Collecting these results, we conclude that the first-order equations (\ref{2.13}) and (\ref{2.14}),  considered as a system for determining of mutually independent dynamical variables $R(t)$ and $\Omega(t)$, are equivalent to the original system (\ref{2.6}). So they can equally be used to study the evolution of a rigid body. 
The initial conditions for the problem (\ref{2.13}) and (\ref{2.14}) are $R_{ij}(0)=\delta_{ij}$, $\Omega_i(0)=\omega_i(0)=(I^{-1}{\bf m})_i$, see (\ref{4.3.0}). 

Any solution $R_{ij}(t)$ that obeys these conditions at $t=0$ authomatically will be orthogonal matrix at any future instant of time. That is we do need to add the constraint $R^TR={\bf 1}$ to the system. To see this, we contract (\ref{2.14}) with $R_{ip}$, obtaining $({\bf R}_p, \dot{\bf R}_j)=-\epsilon_{pjk}\Omega_k$. This implies $(d/dt)({\bf R}_p, {\bf R}_j)=0$, or $({\bf R}_p, {\bf R}_j)=\mbox{const}$. We conclude that $({\bf R}_p(t), {\bf R}_j(t))=({\bf R}_p(0), {\bf R}_j(0))=\delta_{ij}$.  Thus, a rigid body can be described using only the differential equations (\ref{2.13}) and (\ref{2.14}). They have the normal form, that is the time derivatives are separated on l.h.s. of the equations. Then the theory of differential equations guarantees the existence and uniqueness of a solution to the Cauchy problem. Note that to prove the existence of solutions for a mixed system of differential and algebraic equations, much more effort is required \cite{AAD_2022}.

Thus, we have achieved our goal of obtaining the equations of motion. Evolution of rotational degrees of freedom can be determined either from the second-order equations (\ref{2.6}) or from the system (\ref{2.13}), (\ref{2.14}). In the latter case, solving Eq. (\ref{2.13}), we obtain the vector of angular velocity in the body $\Omega_i(t)$. With this $\Omega_i(t)$, we should solve Eq. (\ref{2.14}) for $R(t)$. Then the dynamics of any point of the rigid body is 
\begin{eqnarray}\label{2.15}
{\bf y}_N(t)={\bf C}_0+{\bf V}_0 t+R(t){\bf x}_N(0). 
\end{eqnarray}
The movement consist of rectilinear motion ${\bf C}_0+{\bf V}_0 t$ of the center of mass, and the ortogonal transformation $R(t){\bf x}_N(0)$ around the center of mass. 

{\bf Conservation of angular momentum and the Euler equations.} \label{SeAM} The equation (\ref{s2.4_1}), that states the conservation of angular momentum, turns out to be equivalent to the Euler equations (\ref{2.13}). 

Indeed, preservation in time of angular momentum (\ref{s2.4_1}) implies precession of angular momentum in the body around the vector of angular velocity in the body
\begin{eqnarray}\label{4.7}
\frac{d{\bf m}}{dt}=\frac{d(R{\bf M})}{dt}=\dot R{\bf M}+R\dot{\bf M}=0, \quad \mbox{then} ~\quad \dot{\bf M}=-R^T\dot R{\bf M}=\hat\Omega{\bf M}, 
\end{eqnarray}
or
\begin{eqnarray}\label{4.8}
\dot{\bf M}=-[{\boldsymbol\Omega}, {\bf M}].
\end{eqnarray}
Substituting  ${\bf M}=I{\boldsymbol\Omega}$ into this equation, we arrive at the Euler equations
\begin{eqnarray}\label{4.8}
I\dot{\boldsymbol\Omega}=[I{\boldsymbol\Omega}, {\boldsymbol\Omega}].
\end{eqnarray}

Let's finish this section with some relevant comments. 

\noindent {\bf 1.} We emphasise that the functions $\Omega_i(t)$ represent componens of instantaneous angular velocity ${\boldsymbol\omega}$ in the body-fixed basis ${\bf R}_i$. Therefore, knowing the solution of the Euler equations, we still cannot say anything definite about the behavior of a rigid body. To do this, it is necessary to solve the equations (\ref{2.14}). 

\noindent {\bf 2.} Computing derivative of $\omega_i(t)=R_{ij}(t)\Omega_j(t)$ and using (\ref{2.14}) we get $\dot\omega_i=R_{ij}\dot\Omega_j$.  So $\dot\omega_i\ne 0$ implies $\dot\Omega_j(t)\ne 0$. This means that if the axis ${\boldsymbol\omega}$ moves in space, it must also move in the body.

\noindent {\bf 3.} As a consequence of (\ref{2.13}) and (\ref{2.14}),  the vector of angular velocity ${\boldsymbol\omega}$ obeys to rather complicated equations of motion 
\begin{eqnarray}\label{3.20.1}
\dot {\boldsymbol\omega}=RI^{-1}R^T[{\bf m}, {\boldsymbol\omega}]. 
\end{eqnarray}
In obtaining of Eq.  (\ref{3.20.1}) we used the identity (\ref{0.3}). 

\noindent {\bf 4.} Replacing ${\boldsymbol\Omega}=R^T{\boldsymbol\omega}$ in Eq. (\ref{2.14}), this turn into (\ref{1.27.1}). 

\noindent {\bf 5.} In Sect. \ref{SeHE} we show that the equations (\ref{2.13}) and (\ref{2.14}) are just the Hamiltonian equations of motion of the theory (\ref{2.1}).

\section{Qualitative picture of motion according to Poinsot.}\label{SePo}

According to the equation $\dot{\bf x}_N=[\boldsymbol{\omega}, {\bf x}_N]$, in the center-of-mass system a rigid body rotates around the axis $\boldsymbol{\omega}$, which in turn moves in space according to Eq. (\ref{3.20.1}). This complicated motion has been visualised by L. Poinsot \cite{Poin}.

We recall that basis vectors in the body, ${\bf R}_j(t)$, were chosen in the direction of the inertia axis. Besides, at the initial instant of time they coincide with basis vectors of the Laboratory, ${\bf R}_j(0)={\bf e}_j$, so at the instant $t$ we have 
\begin{eqnarray}\label{5.1}
{\bf R}_j(t)={\bf e}_i R_{ij}(t).
\end{eqnarray}   
Denote coordinates of the radius-vector of a spatial point ${\bf x}$ in the laboratory system  by $x_i$, while in the body system by $z_i(t)$.  The identity ${\bf x}={\bf e}_ix_i={\bf R}_i(t)z_i(t)$ together with (\ref{5.1}) implies ${\bf e}_ix_i={\bf e}_i R_{ij}(t)z_j(t)$, then the coordinates are related as follows:
\begin{eqnarray}\label{5.2}
x_i=R_{ij}z_j, \qquad z_i=R^T_{ij}x_j. 
\end{eqnarray}
If ${\bf x}(t)$ represents the trajectory of some particle moving in the space, derivative of these equalities gives the relation between velocity vectors in the laboratory and in the body systems 
\begin{eqnarray}\label{5.3}
\dot x_i-[\boldsymbol{\omega}, {\bf x}]_i=R_{ij}\dot z_j, \qquad \dot z_i+\epsilon_{ijk}\Omega_j z_k=R^T_{ij}\dot x_j, 
\end{eqnarray}
where $\Omega_i$ are coordinates of angular velocity $\boldsymbol{\omega}$ in the body. 
If ${\bf x}(t)$ is a point of the moving body, then $\dot z_i=0$, and the equation (\ref{5.3}) implies $\dot{\bf x}=[\boldsymbol{\omega}, {\bf x}]$, as it should be. If ${\bf x}$ is a spatial point, then $\dot x_i=0$, and its velocity in the body system is $\dot z_i=-\epsilon_{ijk}\Omega_j z_k$. 

\begin{figure}[t] \centering
\includegraphics[width=10cm]{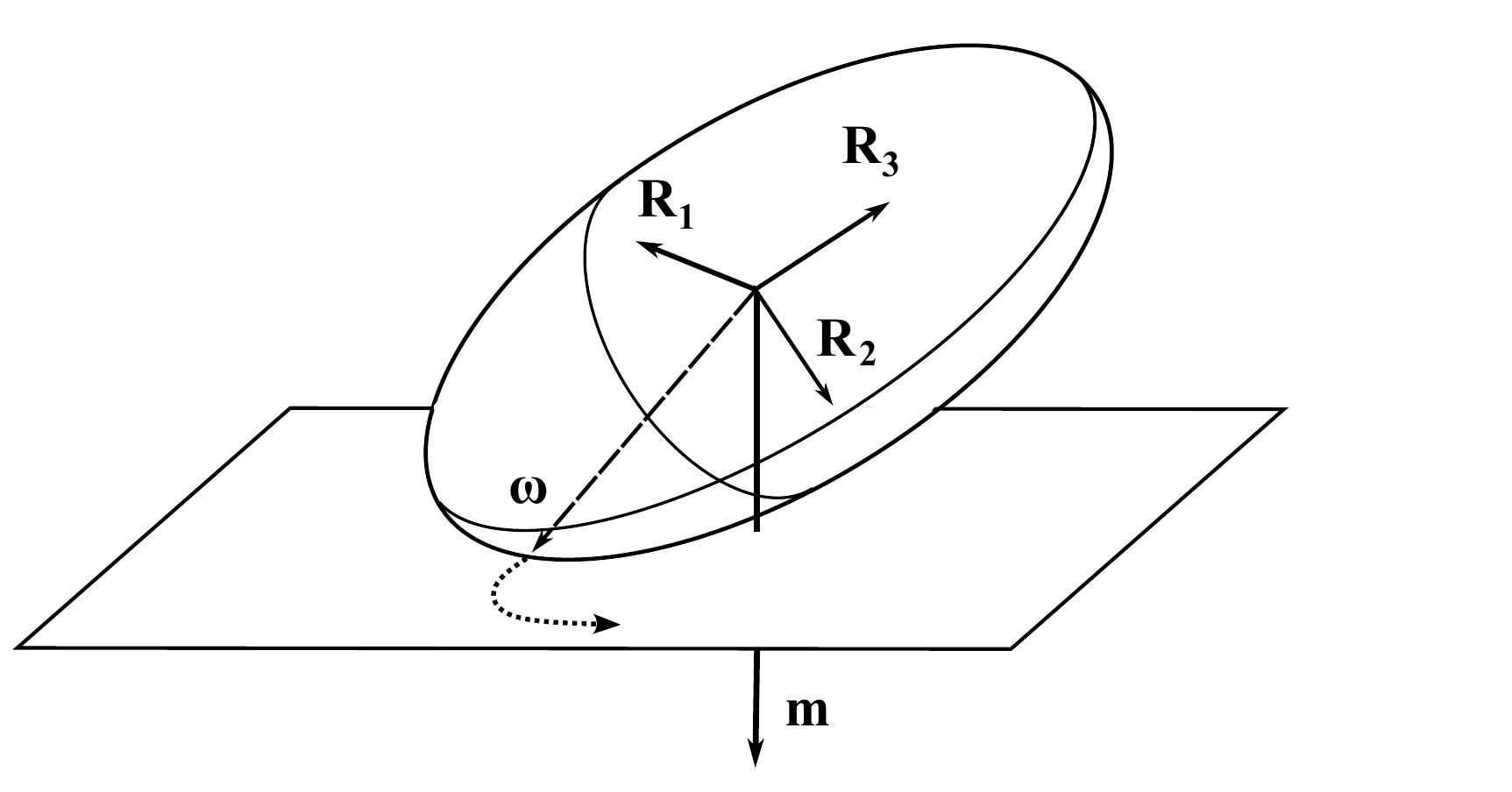}
\caption{Poinsot's ellipsoid rolls on the invariable plane without slipping.}\label{TvT_5}
\end{figure}

Consider the motion of a rigid body with given angular momentum ${\bf m}=\mbox{const}$ and with the 
energy $E=\frac12\sum_iI_i^{-1}m_i^2$. The expressions for conserved energy (\ref{1.31}) prompt to associate with the body at each instant $t$ the ellipsoid with axes in the direction of the body axes ${\bf R}_i(t)$, and with the values of semiaxes equal to $\sqrt{2E/I_i}$. By construction, in the coordinate system ${\bf R}_i(t)$ its equation is of canonical form   
\begin{eqnarray}\label{5.4}
\frac12 I_1z_1^2+\frac12 I_2z_2^2+\frac12 I_3z_3^2=E. 
\end{eqnarray}
The conservation of energy in the form $\frac12\sum I_i\Omega_i^2(t)=E$ implies, that the functions $\Omega_i(t)$ obey to this equation, that is the end of raduis-vector of angular velocity $\boldsymbol{\omega}(t)$ always lies on the ellipsoid.  Using Eq. (\ref{5.2}), the equation of ellipsoid in the laboratory system is 
\begin{eqnarray}\label{5.5}
f(x_1, x_2, x_3)\equiv  \frac12 (R(t)IR^T(t))_{ij}x_i x_j-E=0. 
\end{eqnarray}
This is called the Poinsot's ellipsoid, see Figure \ref{TvT_5}. 
The conservation of energy in the form $\frac12\sum (RIR^T)_{ij}\omega_i\omega_j=E$ implies that the functions $\omega_i(t)$ obey to this equation. That is, once again, the end of raduis-vector of angular velocity $\boldsymbol{\omega}(t)$ always lies on the moving ellipsoid. 

Since the axes of the ellipsoid coincide with coordinate axes ${\bf R}_i(t)$ of the moving body, the position of the Poinsot's ellipsoid in space at each instant of time visualizes also the position of the body itself. 

For the latter use, we compute gradient of the function $f$.  This gives the following expression for the normal vector to the ellipsoid surface: $\overrightarrow{\mbox{grad} ~f}_i=(RIR^T)_{ij}x_j$. At the point of ellipsoid $\boldsymbol{\omega}(t)$, direction of the normal vector coincide with the direction of constant vector ${\bf m}$ (see Eq. (\ref{2.8}) )
\begin{eqnarray}\label{5.6}
\overrightarrow{\mbox{grad} ~ f}(\boldsymbol{\omega}(t))=R(t)IR^T(t)\boldsymbol{\omega}(t)={\bf m}. 
\end{eqnarray}

The conservation of energy in the form $\frac12(\boldsymbol{\omega}(t), {\bf m})=E$ implies, that projection of angular velocity $\boldsymbol{\omega}(t)$ on the direction of angular momentum ${\bf m}$ is the same number at each instant of time 
\begin{eqnarray}\label{5.7}
|\boldsymbol{\omega}(t)|\cos\alpha(t)=\frac{2E}{|{\bf m}|}=\mbox{const}, 
\end{eqnarray}
that is radius-vector $\boldsymbol{\omega}(t)$ moves on the plane that is orhogonal to the constant vector ${\bf m}$ and lies at the 
distance $\frac{2E}{|{\bf m}|}$ from the center of coordinate system. This is called the {\it invariable plane}. 

According to (\ref{5.6}), at the point $\boldsymbol{\omega}$ the normal vector $\overrightarrow{\mbox{grad} ~ f}(\boldsymbol{\omega}(t))$  to the Poinsot's ellipsoid is othogonal to the invariable plane, so it touches the plane without crossing it. The end of the 
vector $\boldsymbol{\omega}(t)$ moves simultaneously on the plane and on the ellipsoid. Its velocity with respect to the plane (that is in the laboratory system) is $\dot\omega_i$ while its velocity with respect to ellipsoid (that is in the body frame) is $\dot\Omega_i$. Using Eq. (\ref{5.3}) we get that the two speeds coincide: $R_{ij}\dot\Omega_j=\dot\omega_i-\epsilon _{ikm}\omega_k\omega_m=\dot\omega_i$,  so $|\dot{\boldsymbol{\omega}}|=|R\dot{\boldsymbol{\Omega}}|=|\dot{\boldsymbol{\Omega}}|$. This implies that during equal intervals of time the point $\boldsymbol{\omega}(t)$ travels the same distance both on the plane and on the ellipsoid, that is the Poinsot's ellipsoid rolls on the  invariable plane without slipping. 

The obtained picture for a rigid body in free motion can be resumed as follows. With a rigid body, considered in the center of mass system, we can associate the invariable plane and the Poinsot's ellipsoid. 
The ellipsoid can be used to visualize the position of the body, since at each instant of time the directions of the ellipsoid axes coincide with directions of basis vectors ${\bf R}_i(t)$ fixed in the body.  During the body's motion, the Poinsot's ellipsoid rolls on the invariable plane without slipping. The radius-vector of angular velocity $\boldsymbol{\omega}(t)$ always ends at the point of contact.

\section{Discussion of equations of motion.}\label{EA}

In this section, we present and discuss various forms of the rigid body equations of motion. 

{\bf Linear system of $3+9$ equations.} Resuming the previous sections, dynamics of rotational degrees of freedom can be studied using first order Euler-Poisson equations  for mutually independent dynamical  variables $R_{ij}(t)$ and $\Omega_i(t)$ 
\begin{eqnarray}
I\dot{\boldsymbol\Omega}=[I{\boldsymbol\Omega}, {\boldsymbol\Omega}], \label{s0} \label{6.7}\\   
\dot R_{ij}=-\epsilon_{jkm}\Omega_k R_{im},  \label{s1} \label{6.8}
\end{eqnarray}
that should be solved with universal initial conditions for the rotation matrix: $R_{ij}(0)=\delta_{ij}$. The initial conditions for $\Omega_i$ can be any three numbers, they represent the initial velocity of rotation of the body. The inertia tensor $I$ is assumed to be diagonal: $I=diagonal (I_1, I_2,  I_3)$.  Given the solution $R(t)$, the evolution of the body's point ${\bf y}(t)$ is restored according to the rule: ${\bf y}(t)={\bf C}_0+{\bf V}_0 t+R(t){\bf x}(0)$, where ${\bf x}(0)$ is the initial position of the point in the center-of-mass system, while  the term ${\bf C}_0+{\bf V}_0 t$ describes the motion of the center of mass with respect to the Laboratory. Recall also that columns $R(t)=({\bf R}_1, {\bf R}_2, {\bf R}_3)$ form an orthonormal basis rigidly connected to the body. The rows $R^T(t)=({\bf G}_1, {\bf G}_2, {\bf G}_3)$ represent the laboratory basis vectors ${\bf e}_i$ in the rigid body basis. For example, the numbers ${\bf G}_1(t)=(R_{11}, R_{12}, R_{13})$ are components of the basis vector ${\bf e}_1$ in the basis ${\bf R}_i(t)$. 

The Euler-Poisson equations admite various  integrals of motion. They are the rotational energy 
\begin{eqnarray}\label{s1.1}
2E=(\boldsymbol{\Omega} I \boldsymbol{\Omega})=I_1\Omega_1^2+I_2\Omega_2^2+I_3\Omega_3^2, \label{6.9}
\end{eqnarray}
three components of angular momentum 
\begin{eqnarray}\label{s1.2}
m_i=(RI\boldsymbol{\Omega})_i=I_1R_{i1}\Omega_1+I_2R_{i2}\Omega_2+I_3R_{i3}\Omega_3,  \label{6.9.1}
\end{eqnarray}
and six more integrals 
\begin{eqnarray}\label{s1.2.0}
R_{ki}R_{kj}=a_{ij}. 
\end{eqnarray}
The integration constants $a_{ij}$ are fixed by the initial conditions: $R_{ij}(0)=\delta_{ij}$ implies $a_{ij}=\delta_{ij}$, so these integrals of motion turn into the ortogonality conditions. Due to this, any  solution $R_{ij}(t)$ to the system with these initial conditions authomatically will be the orthogonal matrix at any future instant: $R^T(t)R(t)=1$.  Besides, the initial conditions imply the following relation between the energy and angular momentum: 
\begin{eqnarray}\label{s1.3}
2E=\frac{1}{I_1}m_1^2+\frac{1}{I_2}m_2^2+\frac{1}{I_3}m_3^2. 
\end{eqnarray}
The square of angular momentum does not contain $R_{ij}$
\begin{eqnarray}\label{s1.4}
{\bf m}^2=I_1^2\Omega_1^2+I_2^2\Omega_2^2+I_3^2\Omega_3^2,
\end{eqnarray}
so the Euler equations itself admite two independent integrals of motion (\ref{s1.1}) and  (\ref{s1.4}). 

The equations (\ref{s1})  are still written for an excess number of variables. Indeed, at each instant of time, the nine matrix elements $R_{ij}$ obey to six constraints $R^TR =1$, so we need to know only some $9-6=3$ independent parameters to specify the matrix $R$. 
It becomes a centenary tradition in the text-books to discuss solutions to these equations using some irreducible set of variables like the Euler angles \cite{Whit_1917,Mac_1936,Lei_1965}. However, there are a number of arguments against this way of presentation.  
First, in the discussions based on the Euler angles, the equations (\ref{s1}) are so dissolved in the calculations that sometimes even not mentioned. 
Second, to describe a rigid body, we need to know namely the evolution of $R_{ij}(t)$. If so, why do we then insist on introducing independent variables? 
Third, the description in terms of independent variables often turns out to be local, which can lead to misunderstandings, see \cite{AAD23_6}. Finally, solving the Euler-Poisson equations  directly for the original variables sometimes requires less effort than solving the same equations through independent variables, see for example the solution to Euler equations for the free asymmetric top in \cite{Landau_8}. So we postpone the introduction of Euler angles until Appendix 1, and here discuss the equations of motion in the original variables.

{\bf Linear system of $3+3$ equations.} Using the lines ${\bf G}_1, {\bf G}_2$ and ${\bf G}_3$ of the matrix $R$, the system (\ref{s1}) reads as follows 
$I\dot{\boldsymbol\Omega}=[I{\boldsymbol\Omega}, {\boldsymbol\Omega}]$,  $\dot{\bf G}_i=[{\bf G}_i, {\boldsymbol\Omega}]$. Therefore, the original system splits into three. It is sufficient to solve only one system of six equations, that consist of three Euler equations and three equations for any one among the vectors ${\bf G}_i$, 
say, ${\boldsymbol\gamma}$:  
\begin{eqnarray}\label{s2}
I\dot{\boldsymbol\Omega}=[I{\boldsymbol\Omega}, {\boldsymbol\Omega}], 
\end{eqnarray}
\begin{eqnarray}\label{s2.1} 
\dot{\boldsymbol\gamma}=[{\boldsymbol\gamma}, {\boldsymbol\Omega}].  
\end{eqnarray}
Then the whole rotation matrix can be restored, choosing three particular solutions to this system: ${\bf G}_1(t)$ is the solution ${\boldsymbol\gamma}(t)$ that obeys the initial condition ${\boldsymbol\gamma}(0)=(1, 0, 0)$, ${\bf G}_2(t)$ is the solution ${\boldsymbol\gamma}(t)$ that obeys the initial condition ${\boldsymbol\gamma}(0)=(0, 1, 0)$, and ${\bf G}_3(t)$ is the solution ${\boldsymbol\gamma}(t)$ that obeys the initial condition ${\boldsymbol\gamma}(0)=(0, 0, 1)$. The Euler equations have two integrals of motion (\ref{s1.1}) and  (\ref{s1.4}). 
The equations (\ref{s2.1}) also have two integrals of  motion following from (\ref{s1.2}) and  (\ref{s1.2.0}): 
\begin{eqnarray}\label{s2.3}
{\boldsymbol\gamma}^2=1, \qquad  c=I_1\Omega_1\gamma_1+I_2\Omega_2\gamma_2+I_3\Omega_3\gamma_3, 
\end{eqnarray}
where $c=m_1$ for ${\boldsymbol\gamma}={\bf G}_1$, and so on.  Using them, the system (\ref{s2}), (\ref{s2.1}) reduces to two first-order equations, each for its own variable. In the case of asymmetric body, they can be formally integrated and give the answer in terms of elliptic 
integrals \cite{Landau_8}.  But in the case of Lagrange (symmetric) top there is an analytic solution for $R_{ij}$ that we present and discuss in the next section.

{\bf Partially integrated system of 9 equations.}
Integrals of motion (\ref{s1.1}) and (\ref{s1.2}), being consequences of equations (\ref{6.7}) and (\ref{6.8}), can be added to them. This gives the equivalent system 
\begin{eqnarray}
I\dot{\boldsymbol\Omega}=[I{\boldsymbol\Omega},{\boldsymbol\Omega}], \label{6.70} \\  
\dot R_{ij}=-\epsilon_{jkp}\Omega_k R_{ip}, \label{6.80} \\
\frac12 \sum_i I_i\Omega_i^2=E=\mbox{const},  \label{6.9} \\
RI{\boldsymbol\Omega}={\bf m}=\mbox{const}.  \label{6.9.1}
\end{eqnarray}
Now the Euler equations (\ref{6.70}) are consequences of (\ref{6.9.1}). Indeed, derivative of (\ref{6.9.1}) reads as follows: $\dot RI{\boldsymbol\Omega}+RI\dot{\boldsymbol\Omega}=0$, then $I\dot{\boldsymbol\Omega}=-(R^T\dot R)I{\boldsymbol\Omega}=\hat\Omega I{\boldsymbol\Omega}=[I{\boldsymbol\Omega}, {\boldsymbol\Omega}]$. 
So we can omit the Euler equations from the system (\ref{6.70})-(\ref{6.9.1}). Further, using (\ref{6.9.1}) in the form 
\begin{eqnarray}\label{6.13}
\Omega_k({\bf m, R})=(I^{-1}R^T{\bf m})_k=\frac{1}{I_k}(m_1R_{1k}+m_2R_{2k}+m_3R_{3k}), 
\end{eqnarray} 
in the equations (\ref{6.80}) and (\ref{6.9}), we reduce our system to the following equations for $R_{ij}(t)$, which contain four integration constants $E$, $m_i$:  
\begin{eqnarray}  
\dot R_{ij}=-\epsilon_{jkp}\Omega_k({\bf m, R}) R_{ip}, \label{6.11} \\
\frac12 \sum_i I_i\Omega_i^2({\bf m, R})=E=\mbox{const}, \label{6.12} 
\end{eqnarray}
Their attraction is that they contain only the rotation matrix, which is what we need to describe a rigid body. However, compared with the previous ones, this system is non-linear in $R_{ij}$.

\section{Examples of solutions in elementary functions.}\label{IN}

{\bf Asymmetrical body with special values of angular momentum.}
Here we consider the equations of asymmetrical body (\ref{6.11})-(\ref{6.12}) with special initial conditions, that admite a separation of variables and then can be solved in analytic form.  It is curious that this solution cannot be obtained using the Euler angles, see Appendix 1 for the details.

\begin{figure}[t] \centering
\includegraphics[width=10cm]{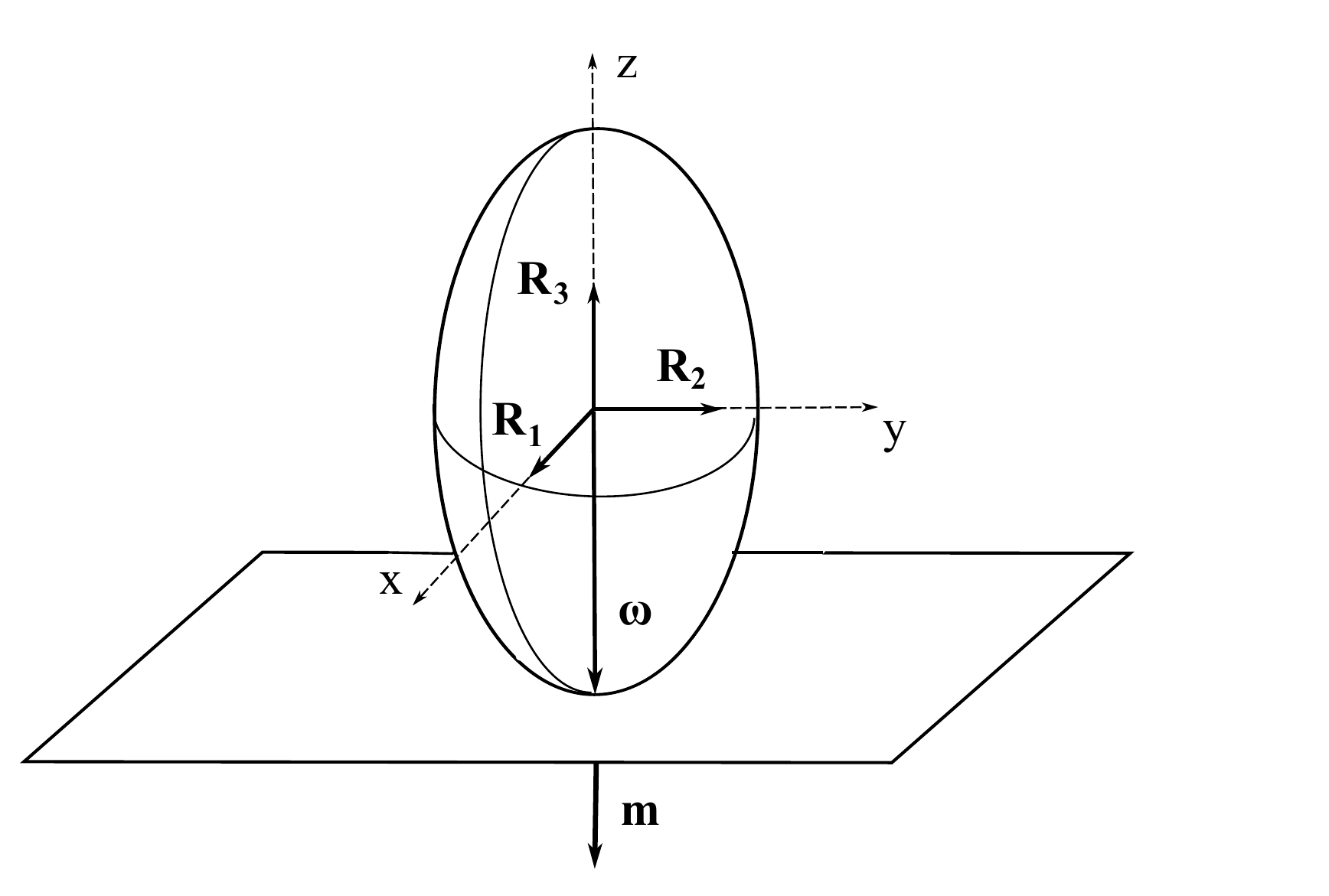}
\caption{Equations of asymmetric body can be easily integrated when angular momentum ${\bf m}$ and the inertia axis ${\bf R}_3$ are collinear at the initial moment.}\label{TvT_7}
\end{figure}

Consider the motion with conserved angular momentum directed along ${\bf e}_3$ -axis: 
\begin{eqnarray}\label{in.1}
{\bf m}=(0, 0, m_3<0). 
\end{eqnarray}
This determines the energy $E=\frac12\sum_i I^{-1}_im_i^2$, the Poinsot's semiaxes $a_i=\sqrt{2E/I_i}$, and the initial angular 
velocity $\omega_i(0)=I^{-1}_im_i$. Then at initial instant of time, axis of inertia $I_3$ is collinear with the angular momentum, see Figure \ref{TvT_7}. 
Eq. (\ref{6.13}) acquires the simple form
\begin{eqnarray}\label{6.13.1}
\Omega_k({\bf m})=\frac{1}{I_k}m_3R_{3k}. 
\end{eqnarray}
Note that we can not achieve this by a suitable rotation of the Laboratory basis.  
Recall that when writing out the equations (\ref{6.7}) and (\ref{6.8}), we assumed that at initial instant the laboratory and rigid body axes were chosen in the direction of inertia axes, see the end of Sect. \ref{AVM}. Due to this, the tensor of inertia in Eq. (\ref{6.13}) is a diagonal matrix, and we deal with rather simple expression (\ref{6.13.1}).   
If we consider a rigid body with an arbitrary angular momentum, and try to rotate the Laboratory system making ${\bf e}_3$ to be collinear with ${\bf m}$, the diagonal matrix $I$ turn into a symmetric matrix $I'$, and we will still be dealing with the complicated angular velocity\footnote{This point was not taken into account in Sect. 69 of \cite{Whit_1917}, where the author said: "In this system the angular momentum of the body about every line which
passes through the fixed point and is fixed in space is constant (40), and consequently the line through the fixed point for which this angular 
momentum has its greatest value is fixed in space. Let this line, which is called the invariable line, be taken as axis OZ, and let OX and OY be two other axes through the fixed point which are perpendicular to OZ and to each other."} $\Omega_k({\bf m})=m_3R_{3i}I'^{-1}_{ik}$.  

With this angular momentum, we substitute the expression (\ref{6.13.1}) into the equations of motion (\ref{6.11}), obtaining
\begin{eqnarray}\label{6.13.01}
\dot R_{ij}=-\omega_k\epsilon_{jkp}R_{3k}R_{ip}, \qquad \mbox{where} \quad \omega_k\equiv\frac{m_3}{I_k}.
\end{eqnarray}
Let us search for a solution of the form (as we saw above, the problem has unique solution with given initial conditions) 
\begin{eqnarray}\label{6.13.02}
{\bf R}_1(t)=\left( 
\begin{array}{c}
x_1(t) \\ y_1(t) \\ 0
\end{array}
\right), \qquad 
{\bf R}_2(t)=\left( 
\begin{array}{c}
x_2(t) \\ y_2(t) \\ 0
\end{array}
\right), \qquad 
{\bf R}_3(t)=\left( 
\begin{array}{c}
0 \\ 0 \\ 1
\end{array}
\right).
\end{eqnarray}
This implies $R_{3k}=\delta_{3k}$. Using this in Eqs. (\ref{6.13.01}) we get $\dot{\bf R}_j=\omega\epsilon_{jp3}{\bf R}_p$, 
or $\dot{\bf R}_1=\omega{\bf R}_2$, $\dot{\bf R}_2=-\omega{\bf R}_1$, as well as the equation $\dot{\bf R}_3=0$, which is identically satisfied by ${\bf R}_3$ written in Eq. (\ref{6.13.02}). Here $\omega=m_3/I_3$. For the components, the equations for ${\bf R}_1$ and ${\bf R}_2$ read
\begin{eqnarray}\label{6.13.03}
\left\{
\begin{array}{c}
\dot x_1=\omega x_2, \\ \dot y_1=\omega y_2;
\end{array}
\right.  \qquad 
\left\{
\begin{array}{c}
\dot x_2=-\omega x_1 , \\ \dot y_2=-\omega y_1. 
\end{array}
\right.  
\end{eqnarray}
Their general solution is as follows
\begin{eqnarray}\label{6.13.04}
\left\{
\begin{array}{c}
x_1=A\cos\omega t+B\sin\omega t,  \\ y_1=C\sin\omega t+D\cos\omega t; 
\end{array} 
\right.  \qquad 
\left\{
\begin{array}{c}
x_2=-A\sin\omega t+B\cos\omega t,  \\ y_2=C\cos\omega t-D\sin\omega t.
\end{array}
\right.  
\end{eqnarray}
Then the initial conditions $R_{ij}(0)=\delta_{ij}$ imply: $A=C=1$, $B=D=0$. In the result we obtained the solution
\begin{eqnarray}\label{6.13.05}
{\bf R}_1=\left( 
\begin{array}{c}
\cos\omega t \\ \sin\omega t \\ 0
\end{array}
\right), \qquad 
{\bf R}_2=\left( 
\begin{array}{c}
-\sin\omega t \\ \cos\omega t \\ 0
\end{array}
\right), \qquad 
{\bf R}_3=\left( 
\begin{array}{c}
0 \\ 0 \\ 1
\end{array}
\right).
\end{eqnarray}
As it should be expected from the Figure \ref{TvT_7}, this states that the vectors ${\bf R}_1$ and ${\bf R}_2$ rotate around the axis $z$ with constant 
angular frequency $\omega=m_3/I_3$. Note also that they obey the equations of precession around the vector $\boldsymbol{\omega}=(0, 0, m_3/I_3)$:  $\dot{\bf R}_1=[\boldsymbol{\omega}, {\bf R}_1]$, $\dot{\bf R}_2=[\boldsymbol{\omega}, {\bf R}_2]$.

{\bf General solution to equations of motion of the free symmetric body.} 
Consider the rigid body with two coinciding moments of inertia. So we take the equations (\ref{s2})-(\ref{s2.3}) with $I_1=I_2$ and with the conserved angular momentum ${\bf m}=(m_1, m_2, m_3)$. First, we confirm that without loss of generality we can assume that $m_1=0$. 

The moments of inertia are eigenvalues of the inertia tensor $I$, with eigenvectors being the body fixed axes at $t=0$: $I{\bf R}_i(0)=I_i{\bf R}_i(0)$.  With $I_1=I_2$ we have $I{\bf R}_1(0)=I_2{\bf R}_1(0)$ and $I{\bf R}_2(0)=I_2{\bf R}_2(0)$, then any linear combination $\alpha{\bf R}_1(0)+\beta{\bf R}_2(0)$ also represents an eigenvector with eigenvalue $I_2$. This means that we are free to choose any two orthogonal axes on the plane ${\bf R}_1(0), {\bf R}_2(0)$ as the inertia axes. We recall that at $t=0$, the Laboratory axes should be chosen along ${\bf R}_i$, since only in this case equations of motion contain $I_i$ instead of the tensor $I_{ij}$. Hence, in the case $I_1=I_2$ we can rotate the Laboratory axes in the plane $(x^1, x^2)$  without breaking the diagonal form of the inertia tensor. Using this freedom, we can assume that $m_1=0$ for our problem. 

{\bf General solution to the Euler equations.}  With $I_1=I_2$, the Euler equations read
\begin{eqnarray}\label{s3}
\dot\Omega_1=\frac{(I_2-I_3)\Omega_3}{I_2}\Omega_2, \qquad \dot\Omega_2=-\frac{(I_2-I_3)\Omega_3}{I_2}\Omega_1, \qquad \dot\Omega_3=0,
\end{eqnarray}
or
\begin{eqnarray}\label{s4}
\dot\Omega_1 =\phi\Omega_2, \qquad \dot\Omega_2=-\phi\Omega_1, \qquad \Omega_3=const,
\end{eqnarray}
where $\phi\equiv(I_2-I_3)\Omega_3/I_2$.  
Their general solution is
$\Omega_1=a\sin(\phi t+\phi_0)$,  $\Omega_2=a\cos(\phi t+\phi_0)$, $\Omega_3=\mbox{const}$.
The solution has a simple meaning: for an observer of the rigid-body frame ${\bf R}_i$, the angular velocity vector ${\boldsymbol\omega}$ precess around ${\bf R}_3$\,-axis with angular frequency $\phi$. 

Let us relate the integration constants $a$, $\Omega_3$ and $\phi_0$ with the basic quantities of a rigid body. The conserved angular 
momentum $m_i$ and the angular velocity $\Omega_i$ are related as follows: $m_i=I_{ij}\Omega_j(0)$, so $m_1=I_2a\sin\phi_0$, $m_2=I_2a\cos\phi_0$,
$m_3=I_3\Omega_3$. Our choice $m_1=0$  fixes the constant $\phi_0=0$. Then $a=m_2/I_2$ and $\Omega_3=m_3/I_3$. In the result, the first oscillation frequency in the problem is determined by third component of conserved angular momentum: $\phi=(I_2-I_3)m_3/I_2 I_3$.   Taking all this into account, the solution is 
\begin{eqnarray}\label{s5}
\Omega_1=\frac{m_2}{I_2}\sin \phi t, \qquad \Omega_2=\frac{m_2}{I_2}\cos\phi t, \qquad \Omega_3=\frac{m_3}{I_3}.
\end{eqnarray}
The explicit form of the solution shows the conservation of the rotational energy: $2E=I_i\Omega_i^2(t)=m_2^2/I_2+m_3^2/I_3$.

{\bf General solution to the Poisson equations.} Take the linear integral of motion from Eq. (\ref{s2.3}) and the third component of the Poisson equation
\begin{eqnarray}\label{s6}
\Omega_2\gamma_1-\Omega_1\gamma_2=\dot\gamma_3, \qquad I_2\Omega_1\gamma_1+I_2\Omega_2\gamma_2=c-I_3\Omega_3\gamma_3.
\end{eqnarray}
The solution to this linear system is
\begin{eqnarray}\label{s7}
\gamma_1=\frac{I_2}{m_2^2}\left[I_2\Omega_2\dot\gamma_3+(c-m_3\gamma_3)\Omega_1\right], \qquad 
\gamma_2=\frac{I_2}{m_2^2}\left[-I_2\Omega_1\dot\gamma_3+(c-m_3\gamma_3)\Omega_2\right].
\end{eqnarray}
Substituting these expressions into the Poisson equation $\dot\gamma_1=\Omega_3\gamma_2-\Omega_2\gamma_3$ we get closed equation for $\gamma_3$
\begin{eqnarray}\label{s8}
\ddot\gamma_3+k^2\gamma_3=\frac{m_3}{I_2^2} c, \qquad 
\mbox{where} \quad k^2\equiv\frac{m_2^2+m_3^2}{I_2^2}=\frac{{\bf m}^2}{I_2^2}, 
\end{eqnarray}
with the general solution 
\begin{eqnarray}\label{s9}
\gamma_3=b\cos(kt+k_0)+\frac{m_3 c}{{\bf m}^2},  \qquad 
\mbox{where} \quad  k\equiv\frac{|{\bf m}|}{I_2}. 
\end{eqnarray}
According to Eq. (\ref{s9}), the second oscillation frequency $k$ in the problem is determined by magnitude of conserved angular momentum. Using
this $\gamma_3$ in equations (\ref{s7}), we get the general solution to the system  (\ref{s2.1}) as follows
\begin{eqnarray}\label{s10}
\gamma_1=-\frac{|{\bf m}|b}{m_2}\cos\phi t\sin(kt+k_0)+\left[\frac{m_2 c}{{\bf m}^2}-\frac{m_3 b}{m_2}\cos(kt+k_0)\right]\sin\phi t, \cr
\gamma_2=\frac{|{\bf m}|b}{m_2}\sin\phi t\sin(kt+k_0)+\left[\frac{m_2 c}{{\bf m}^2}-\frac{m_3 b}{m_2}\cos(kt+k_0)\right]\cos\phi t, \cr 
\gamma_3=b\cos(kt+k_0)+\frac{m_3 c}{{\bf m}^2}. \qquad \qquad \qquad \qquad \qquad \qquad \qquad \quad 
\end{eqnarray}
At the initial instant $t=0$ we get
\begin{eqnarray}\label{s11}
\gamma_1(0)=-\frac{|{\bf m}|b}{m_2}\sin k_0,  \quad
\gamma_2(0)=\frac{m_2 c}{{\bf m}^2}-\frac{m_3 b}{m_2}\cos k_0, \quad
\gamma_3(0)=b\cos k_0 +\frac{m_3 c}{{\bf m}^2}. 
\end{eqnarray}
As we saw above, the three rows of the matrix $R_{ij}$ are obtained from Eqs. (\ref{s10}) if we choose the integration constants $b$ and $k_0$ so 
that ${\boldsymbol\gamma}(0)=(1, 0, 0)$ with $c=m_1$, then ${\boldsymbol\gamma}(0)=(0, 1, 0)$ with $c=m_2$, and at last ${\boldsymbol\gamma}(0)=(0, 0, 1)$ with $c=m_3$. Solving the equations (\ref{s11}) with these dates, we get, in each case 
\begin{eqnarray}\label{s12}
c=m_1=0, \quad b=-\frac{m_2}{|{\bf m}|}, \quad k_0=\frac{\pi}{2}; \quad
c=m_2, \quad b=-\frac{m_2 m_3}{{\bf m}^2}, \quad k_0=0; \quad
c=m_3, \quad b=\frac{m_2^2}{{\bf m}^2}, \quad k_0=0. 
\end{eqnarray}
Substituting these values into Eq. (\ref{s10})  we get final form of the rotation matrix $R$ of the free symmetric top as follows:
\begin{eqnarray}\label{s14}
\left(
\begin{array}{ccc}\label{s12.11}
\cos kt\cos\phi t-\hat m_3\sin kt\sin\phi t & -\cos kt\sin\phi t-\hat m_3\sin kt\cos\phi t  &  \hat m_2\sin kt  \\
{} & {} & {} \\
\hat m_3 \sin kt\cos\phi t+(\hat m_2^2 +\hat m_3^2\cos kt)\sin\phi t & 
-\hat m_3 \sin kt\sin\phi t+(\hat m_2^2 +\hat m_3^2\cos kt)\cos\phi t & \hat m_2\hat m_3(1-\cos kt) \\
{} & {} & {} \\
-\hat m_2 \sin kt\cos\phi t+\hat m_2 \hat m_3(1-\cos kt)\sin\phi t &
\hat m_2 \sin kt\sin\phi t+\hat m_2 \hat m_3(1-\cos kt)\cos\phi t & \hat m_3^2 +\hat m_2^2\cos kt
\end{array}\right) 
\end{eqnarray}
where, assuming $|{\bf m}|\ne 0$, we denoted  by $\hat m_i=m_i/|{\bf m}|$ the components of unit vector in the direction of conserved angular momentum. The two frequences in the problem are $\phi=\frac{I_2-I_3}{I_2 I_3}m_3$ and $k=|{\bf m}|/I_2=\sqrt{m_2^2+m_3^2}/I_2$. Curiously enough, the matrix $R$ depends on the moments of inertia only through the frequences. 

Formulas (\ref{2.15}) and (\ref{s12.11}) solves the problem of motion of the free symmetric top.

{\bf Decomposition of the rotation matrix on two subsequent rotations, and Poinsot's picture of motion.} 
We begin our discussion of the rotation matrix by looking at various limiting cases. 

{\bf (1)} Consider the conserved angular momentum ${\bf m}$ in the direction of laboratory axis $OY$, that is $m_3=0$, $m_2\ne 0$. This 
implies $\phi=0$, $|{\bf m}|=|m_2|$, $\hat m_2=\pm 1$, and $\hat m_2\sin |{\bf m}|t/I_2=\sin m_2 t/I_2$. Using this in Eq. (\ref{s14}) we get
\begin{eqnarray}\label{s15}
R_{OY}(t)=\left(
\begin{array}{ccc}
\cos \frac{m_2t}{I_2} & 0 & \sin \frac{m_2t}{I_2} \\ 
0 & 1 & 0 \\
-\sin \frac{m_2t}{I_2} & 0 & \cos \frac{m_2t}{I_2} 
\end{array}\right).  
\end{eqnarray}
This single-frequency motion represents the rotation of the rigid body around the inertia axis $OY$. 

{\bf (2)} Similarly, taking $m_2=0$, $m_3\ne 0$ we get $|{\bf m}|=|m_3|$, $\hat m_3=\pm 1$, $k=|m_3|/I_2$ and $\phi+m_3/I_2=m_3/I_3$. 
Using this in 
Eq. (\ref{s14}) we observe that two frequences are combined into one as follows
\begin{eqnarray}\label{s16}
R_{OZ}(t)=\left(
\begin{array}{ccc}
\cos \frac{m_3t}{I_3}  & -\sin \frac{m_3t}{I_3} & 0  \\ 
\sin \frac{m_3t}{I_3} &  \cos \frac{m_3t}{I_3} & 0 \\
0 & 0 & 1 \\
\end{array}\right).  
\end{eqnarray}
This single-frequency motion represents the rotation of the rigid body around the inertia axis $OZ$. Note that it coincides with the motion of an asymmetric top (\ref{6.13.05}) with the same initial conditions. 

{\bf (3)} For the totally symmetric body $I_1=I_2=I_3$ we get $\phi=0$, and the rotation matrix (\ref{s14}) acquires the following form: 
\begin{eqnarray}\label{s17}
R_{\bf m}(t)=\left(
\begin{array}{ccc}
\cos kt & -\hat m_3\sin kt  &  \hat m_2\sin kt  \\
{} & {} & {} \\
\hat m_3 \sin kt & 
\hat m_2^2 +\hat m_3^2\cos kt & \hat m_2\hat m_3(1-\cos kt) \\
{} & {} & {} \\
-\hat m_2 \sin kt & \hat m_2 \hat m_3(1-\cos kt) & \hat m_3^2 +\hat m_2^2\cos kt
\end{array}\right). 
\end{eqnarray}
where $k=|{\bf m}|/I_2$.  The points of the body lying along the axis ${\bf m}$ remain at rest during this movement: for any $c\in {\mathbb R}$ we get $R_{\bf m}(t)c{\bf m}=c{\bf m}$. So this motion represents the rotation of the body around the conserved angular momentum vector ${\bf m}$. This is the only possible motion of totally symmetric body. 

\begin{figure}[t] \centering
\includegraphics[width=12cm]{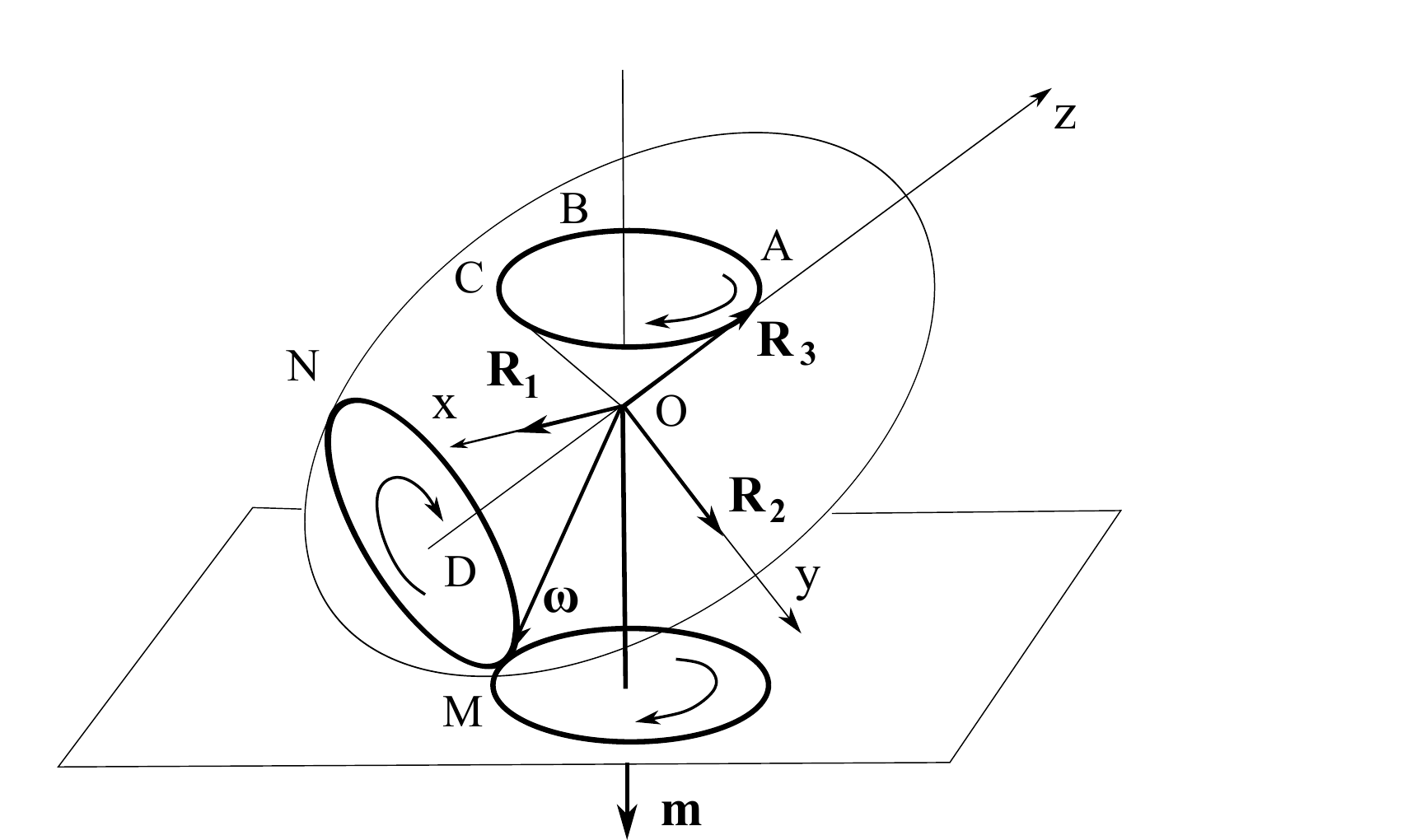}
\caption{Poinsot's picture for the symmetric top.}\label{TvT_8}
\end{figure}

{\bf (4)} To discuss the general case of two-frequency motion (\ref{s14}), we recall how the rotational motions of the rigid body look like in the picture of Poinsot.  Let's put 
\begin{eqnarray}\label{s18}
I_1=I_2> I_3, \qquad  m_1=0, \qquad m_2>0, \qquad m_3<0. 
\end{eqnarray}
This determines the energy $E=\frac12\sum_i I^{-1}_im_i^2$, the semiaxes $a_i=\sqrt{2E/I_i}$, and the initial angular 
velocity $\omega_i(0)=I^{-1}_im_i$. With these dates, we can construct the invariable plane and the Poinsot's ellipsoid. 
Position of the Poinsot's elipsoid at the initial moment $t=0$ is shown in the Figure \ref{TvT_8}. 
At this moment the vectors ${\bf m}$, ${\boldsymbol\omega}$,  ${\bf R}_2$, ${\bf R}_3$ as well as the points $O$, $A$, $C$, $D$, $M$ and $N$ lie on the plane $x=0$. The point $A$ represents the end point of the body fixed vector ${\bf R}_3$ at this moment. As $\dot{\bf R}_3=[ {\boldsymbol\omega}, {\bf R}_3]$, the point $A$ of the body starts its motion in the direction of the arrow drawn near this point. This determines the directions of motions of other elements as follows. The wheel $D$, with the axis $DO$ fixed at the point $O$, rolls without slipping along the circle drawn on invariable plane. The end of ${\boldsymbol\omega}$ moves on this circle counterclockwise around ${\bf m}$. In the result, all points of the body fixed axis $Dz$ move on the circles of the cone with the axis along the vector ${\bf m}$. In particular, the point $A$ of the body moves on the circle $ACBA$.  

In resume, the axis $Dz$, consisting of the points of the body, generates the surface of the cone, while the remaining points of the body 
instantaneously rotate around this axis. 

The rotation matrix (\ref{s14}) is in correspondence with this picture. Indeed, we observe that it can be decomposed as 
follows: $R(t)=R_{\bf m}(t)\times R_{OZ}(t)$, where $R_{\bf m}(t)$ is the rotation matrix (\ref{s17}), and 
\begin{eqnarray}\label{s19}
R_{OZ}(t)=\left(
\begin{array}{ccc}
\cos \phi t  & -\sin \phi t & 0  \\ 
\sin \phi t &  \cos \phi t & 0 \\
0 & 0 & 1 \\
\end{array}\right), 
\end{eqnarray}
where $\phi=\frac{(I_2-I_3)m_3t}{I_2I_3}$. Then the position ${\bf x}(t)$ of any point of the body at the 
instant $t$ is: ${\bf x}(t)=R_{\bf m}(t)\times R_{OZ}(t){\bf x}(0)$. It is obtained by rotating the initial position vector ${\bf x}(0)$ first around the laboratory axis $OZ$ by the angle $\frac{(I_2-I_3)m_3t}{I_2I_3}$ and then around the ${\bf m}$\,-axis by the angle $\frac{|{\bf m}|t}{I_2}$. 

For completeness, we also present the explicit form for the vector of instantaneous angular velocity of the symmetric top
\begin{eqnarray}\label{s20}
{\boldsymbol\omega}(t)=R(t){\boldsymbol\Omega}(t)=(\omega_1, \omega_2, \omega_3)= \qquad \qquad \qquad \qquad \qquad \qquad \qquad \qquad \qquad \qquad  \cr \left( \frac{(I_2-I_3)m_2 m_3}{I_2I_3|{\bf m}|}\sin kt, ~ 
\frac{(I_3m_2^2+I_2m_3^2)m_2}{I_2I_3{\bf m}^2}-\frac{(I_2-I_3)m_2 m_3^2}{I_2I_3{\bf m}^2}\cos kt, ~ 
\frac{(I_3m_2^2+I_2m_3^2)m_3}{I_2I_3{\bf m}^2}+\frac{(I_2-I_3)m_2^2 m_3}{I_2I_3{\bf m}^2}\cos kt \right)
\end{eqnarray}
The magnitude of this vector gives total frequncy of rotation of the body's points at each instant of 
time: ${\boldsymbol\omega}^2=\frac{m_2^2}{I_2^2}+\frac{m_3^2}{I_3^2}$. It is related with two frequences of the rotation matrix as follows: ${\boldsymbol\omega}^2=k^2+\frac{I_2+I_3}{I_2I_3}m_3\phi$.

\section{Hamiltonian equations for rotational degrees of freedom and the Euler-Poisson equations.}\label{SeHE}

From point of view of classical mechanics, the Euler-Poisson equations of a rigid body (\ref{2.13}) and (\ref{2.14}) are not just a first-order system, but precisely the Hamiltonian system. That is they have the following structure: $z=\{ z, H(z) \}$. Our aim will be to find the explicit form of the Hamiltonian $H(z)$ and the brackets $\{ {}, {} \}$, see Eqs. (\ref{11.26})-(\ref{11.270}) below.  To achieve this, we use the Dirac's version \cite{Dir_1950, GT, deriglazov2010classical} of Hamiltonian formalism, which is well adapted for the analysis of a theory with constraints. 

Introducing the conjugate momenta for all dynamical variables: $p_{ij}=\partial L/\partial\dot R_{ij}$ and $\pi_{ij}=\partial L/\partial\dot\lambda_{ij}$, we obtain the expression for $p_{ij}$ in terms of velocities
\begin{eqnarray}\label{3.2}
p_{ij}=\dot R_{ik}g_{kj}, \quad \mbox{then} ~ \dot R_{ij}=p_{ik}g^{-1}_{kj},
\end{eqnarray}
and the equalities
\begin{eqnarray}\label{3.3}
\pi_{ij}=0,
\end{eqnarray}
called primary constraints of the Dirac formalism. According to  Dirac,  the Hamiltonian of the theory with constraints should contain the primary constraints as follows: $H=p_{ij}\dot R_{ij}-L+\varphi_{ij}\pi_{ij}$, where we should use Eq. (\ref{3.2}) to represent the velocities $\dot R$ through the momenta $p$. The Hamiltonian involves products  of primary constraints with auxiliary variables $\varphi_{ij}(t)$, they also known as the Lagrangian multipliers for the primary constraints. We get
\begin{eqnarray}\label{3.4}
H=\frac12 g^{-1}_{ij}p_{ki}p_{kj}+\frac12 \lambda_{ij}[R_{ki}R_{kj}-\delta_{ij}]+\varphi_{ij}\pi_{ij}. 
\end{eqnarray}
Then the Hamiltonian equations can be obtained with use of Poisson brackets according to the standard rule: $\dot z=\{z, H\}$, where $z$ is any of phase-space variables. The non vanishing fundamental brackets are (there is no summation over $i$ and $j$): $\{ R_{ij}, p_{ij}\}=1$, $\{\lambda_{ij}, \pi_{ij}\}=1$. We obtain 
\begin{eqnarray}\label{3.5}
\dot R=pg^{-1}, \qquad \dot p=-R\lambda,  ~ \cr
\dot\lambda=\varphi, \qquad \dot\pi=R^TR-{\bf 1}. 
\end{eqnarray}
Together wuth the  constraints (\ref{3.3}), these equations  imply $R^TR={\bf 1}$. This algebraic equation, appeared due to the primary constraints, is called the second-stage constraint of the Dirac's formalism. 
It is known \cite{Dir_1950, GT, deriglazov2010classical}, that the first-order equations (\ref{3.5}) together with the constraints (\ref{3.3}) are equivalent to the Lagrangian equations (\ref{1.37}). 

Our Hamiltonian  system consist of algebraic equations $\pi_{ij}=0$ and the first-order equations (\ref{3.5}),  so it can imply the nontrivial self consistency conditions.  They are obtained by calculating the time derivative of the constraints and eliminating from the resulting expressions all the variables with derivative using the first-order equations. To illustrate how this works, consider the theory with the phase-space variables denoted $z^A$, with one original constraint $g(z^A)=0$, and with one auxiliary variable $\varphi$.  Then the Hamiltonian is of the form $H(z^A, \varphi)=H_0(z^A)+\varphi g(z^A)$. The constraint $g(z^A)=0$ is satisfied at all instants $t$, so it implies $\dot g(z^A)=0$. We get: $\dot g(z^A)=\partial_A g(z^A)\dot z^A=\partial_A g\{z^A, H(z, \varphi)\}=\{g(z^A), H(z^A, \varphi)\}=\{g(z^A), H_0(z^A)\}=0$. If the algebraic consequence  $\{g(z^A), H_0(z^A)\}=0$ of the original system is functionally independent from the original constraints, we discovered a new algebraic equation  that must be satisfied by all solutions $ z^A(t)$ of our theory, that is a new constraint.  By repeating the procedure for this constraint, we get either one more new constraint, or an equation that determines $\varphi$, or a constraint that will not be functionally independent from the original constraint.

This, in essence, is the Dirac method \cite{Dir_1950, GT, deriglazov2010classical} for revealing the algebraic equations that may be hidden in the mixed system of  algebraic and first-order equations. As we will see, for our case this procedure allow us to exclude algebraically all the auxiliary variables from the final equations of motion. 

Having carried out this calculation for the constraints $\pi_{ij}=0$, we just get the constraint. Checking its preservation in time,  we obtain six new algebraic equations, called third-stage constraints
\begin{eqnarray}\label{3.6}
\frac{d}{dt}[R_{ki}R_{kj}-\delta_{ij}]=\{R_{ki}R_{kj}, H\}=(R^Tpg^{-1})_{ij}+(i\leftrightarrow j)=0. 
\end{eqnarray}
Denoting $(R^Tpg^{-1})_{ij}={\mathbb P}_{ij}$, we can decompose this matrix on symmetric and antisymmetric parts, ${\mathbb P}_{ij}={\mathbb P}_{(ij)}-\hat\Omega_{ij}$, where
\begin{eqnarray}\label{3.7}
{\mathbb P}_{(ij)}=\frac12[{\mathbb P}_{ij}+{\mathbb P}_{ji}]=\frac12[R^Tpg^{-1}+(R^Tpg^{-1})^T]_{ij}, \qquad 
\hat\Omega_{ij}=-\frac12[{\mathbb P}_{ij}-{\mathbb P}_{ji}]=-\frac12[R^Tpg^{-1}-(R^Tpg^{-1})^T]_{ij},
\end{eqnarray}
Then the constraints (\ref{3.6}) state that symmetric part of this matrix vanishes, $\mathbb P_{(ij)}=0$. We can rewrite these constraints back in the Lagrangian form, substituting $p=\dot Rg$. We get $\mathbb P_{(ij)}=(R^T\dot R)_{ij}+(R^T\dot R)^T_{ij}=0$, which is just a consequence of  the constraint $R^TR={\bf 1}$. The Lagrangian form of antisymmetric part   
\begin{eqnarray}\label{3.8}
\left.-\frac12({\mathbb P}_{ij}-{\mathbb P}_{ji})\right|_{p=\dot Rg}=-(R^T\dot R)_{ij}=\hat\Omega_{ij}=\epsilon_{ijk}\Omega_k,
\end{eqnarray}
is just the angular velocity in the body, see Eq. (\ref{1.29}). On this reason the phase-space quantity $-\frac12({\mathbb P}_{ij}-{\mathbb P}_{ji})$ was denoted by $\hat\Omega_{ij}$.

Testing the preservation with time of the third-stage constraints we get 
\begin{eqnarray}\label{3.9}
\dot{\mathbb P}_{(ij)}=\{{\mathbb P}_{(ij)}, H\}=\frac12 g^{-1}_{ik}p_{nk}\{R_{nj}, p_{mr}p_{ms}\}g^{-1}_{rs}+ \cr 
\frac12 \lambda_{rs}g^{-1}_{ik} R_{nj}\{p_{nk}, R_{mr}R_{ms}\}+(i\leftrightarrow j)=(g^{-1}p^Tpg^{-1}-g^{-1}\lambda)_{ij}+(i\leftrightarrow j)=0. 
\end{eqnarray}
This leads to the following fourth-stage constraints
\begin{eqnarray}\label{3.10}
g^{-1}\lambda+\lambda g^{-1}=2g^{-1}p^Tpg^{-1}, \qquad \mbox{or} ~ ~ (g^{-1}\lambda+\lambda g^{-1})_{ij}=2[{\boldsymbol\Omega}^2\delta_{ij}-\Omega_i\Omega_j].
\end{eqnarray}
To obtain these equalities, were used the constraints (\ref{3.6}). Preservation with time of the fourth-stage constraints gives the algebraic equations that unambiguously determine the auxiliary variables $\varphi$. We do not write them, as they do not 
contribute \cite{deriglazov2010classical} into the equations of motion for dynamical variables $R$ and ${\boldsymbol\Omega}$. 

Let us resume the obtained equations of motion. As the variables $\lambda$ and $\pi$ will be determined from the constraints, the essential dynamical equations are
\begin{eqnarray}\label{3.11}
\dot R=pg^{-1}, \qquad \dot p=-R\lambda.  
\end{eqnarray}
They are accompanied by the chain of constraints
\begin{eqnarray}\label{3.12}
\pi_{ij}=0, \qquad R^TR={\bf 1}, \qquad (R^Tpg^{-1})_{ij}+(i\leftrightarrow j)=0, \qquad g^{-1}\lambda+\lambda g^{-1}=2g^{-1}p^Tpg^{-1}. 
\end{eqnarray}
The equations (\ref{3.7}) and  (\ref{3.8}) prompts to make a change of variables such, that ${\mathbb P}_{(ij)}$ become a part of new coordinates, and the surface of constraints will be then described by trivial equations ${\mathbb P}_{(ij)}=0$. We introduce them as follows: 
\begin{eqnarray}\label{3.13}
(R_{ij}, p_{ij}) \Leftrightarrow (R_{ij}, {\mathbb P}_{(ij)}, \hat\Omega_{ij})  \Leftrightarrow (R_{ij}, {\mathbb P}_{(ij)}, \Omega_k).
\end{eqnarray}
In these variables, we have the trivial constraints $\pi_{ij}=0$, ${\mathbb P}_{(ij)}=0$, while the remaining equations read
\begin{eqnarray}\label{3.14}
\dot R_{ij}=-\epsilon_{jkm}\Omega_k R_{im}, \qquad \dot\Omega_i=-\frac12\epsilon_{ijk}(g^{-1}\lambda)_{jk},  \qquad R^TR={\bf 1}, 
\end{eqnarray}
\begin{eqnarray}\label{3.15}
(g^{-1}\lambda+\lambda g^{-1})_{ij}=2[{\boldsymbol\Omega}^2\delta_{ij}-\Omega_i\Omega_j]. 
\end{eqnarray}
The last equation unambiguously determines $\lambda$ as follows:
\begin{eqnarray}\label{3.16}
\lambda_{ij}=\frac{2g_ig_j}{g_i+g_j}[{\boldsymbol\Omega}^2\delta_{ij}-\Omega_i\Omega_j]. 
\end{eqnarray}
There is no of summation over $i$ and $j$, and $g_i$ are diagonal elements of the mass matrix. This allows us to eliminate $\lambda$ from dynamical equations for ${\boldsymbol\Omega}$, which gives the Euler equations
\begin{eqnarray}\label{3.17}
\dot\Omega_1=-\frac{g_2-g_3}{g_2+g_3}\Omega_2\Omega_3=\frac{1}{I_1}(I_2-I_3)\Omega_2\Omega_3, \cr
\dot\Omega_2=-\frac{g_3-g_1}{g_3+g_1}\Omega_1\Omega_3=\frac{1}{I_2}(I_3-I_1)\Omega_1\Omega_3, \cr
\dot\Omega_3=-\frac{g_1-g_2}{g_1+g_2}\Omega_1\Omega_2=\frac{1}{I_3}(I_1-I_2)\Omega_1\Omega_2,
\end{eqnarray}
where $I_i$ are components of the tensor of inertia, see (\ref{1.35.1}).  In a more compact form, with use of vector product they read
$I\dot{\boldsymbol\Omega}=[I{\boldsymbol\Omega},{\boldsymbol\Omega}]$.  
Note also that $d(R^TR)/dt=0$ is a consequence of the first equation of the system (\ref{3.14}). So, if the condition $R^TR={\bf 1}$ is satisfied at the initial instant of time, it will be authomatically satisfied for any solution to the system (\ref{3.14}) at any instant. Hence, we can solve dynamical equations of this system without worrying about the algebraic constraint. 
In  the result, the Hamiltonian equations of motion\footnote{The Euler equations (\ref{3.19}), considered by themselves without reference to (\ref{3.18}), form the Hamiltonian system with a degenerate Poisson structure, see \cite{Ner_2022}.} for rotational degrees of freedom are
\begin{eqnarray}\label{3.18} 
\dot R_{ij}=-\epsilon_{jkm}\Omega_k R_{im}, 
\end{eqnarray}
\begin{eqnarray}\label{3.19}
I\dot{\boldsymbol\Omega}=[I{\boldsymbol\Omega}, {\boldsymbol\Omega}].   
\end{eqnarray}
They coincide with the first-order equations (\ref{2.13}) and (\ref{2.14}), that were obtained in Sect. \ref{FOE}.

\section {Chetaev bracket is the Dirac bracket.}\label{Chet}
In the previous section we demostrated that the Euler-Poisson equations (\ref{3.18}), (\ref{3.19})  are just the Hamiltonian equations of the theory (\ref{2.1}). However, the final form of Hamiltonian and of Poisson brackets for these equations has not been presented. We will get them in this section, for which we partially repeat the previous calculations in a more formal, but faster way. To this aim, it is convenient first to rewrite the original variational problem (\ref{2.1}) in terms of the angular velocity in the body
\begin{eqnarray}\label{11.1}
M_i(R, \dot R)=I_{ik}\Omega_k=-\frac12 I_{ik}\epsilon_{knm}(R^T\dot R)_{nm}. 
\end{eqnarray}
The Lagrangian (\ref{2.1}) acquires the form\footnote{The two variational problems are equivalent, see the Appendix 2.} 
\begin{eqnarray}\label{11.2}
L=\frac12 I^{-1}_{ij}M_i M_j-\frac12\lambda_{ij}[R_{ki}R_{kj}-\delta_{ij}]. 
\end{eqnarray}
Introducing the conjugate momenta for all dynamical variables: $\pi_{ij}=\partial L/\partial\dot\lambda_{ij}$ and $p_{ij}=\partial L/\partial\dot R_{ij}$, we obtain the primary constraints $\pi_{ij}=0$ and the following expression for $p_{ij}$ in terms of velocities: 
\begin{eqnarray}\label{11.3}
p_{ij}=-\frac12 R_{in}\epsilon_{njk}M_k(R, \dot R), \qquad \mbox{or} \quad (\tilde Rp)_{ij}=-\frac12\epsilon_{ijk}M_k(R, \dot R), \quad \mbox{where} \quad \tilde R\equiv R^{-1}. 
\end{eqnarray}
It implies six more primary constraints
\begin{eqnarray}\label{11.4}
P_{ij}\equiv[(\tilde Rp)+(\tilde Rp)^T]_{ij}=0,
\end{eqnarray}
and the expression for $M_i$ (note that $R$ is for now an arbitrary matrix)
\begin{eqnarray}\label{11.5}
M_i(R, \dot R)=-\epsilon_{ijk}(\tilde Rp)_{jk}, \qquad \mbox{then} \quad [(R^T\dot R)-(R^T\dot R)^T]_{ij}=2\epsilon_{ijk}I^{-1}_{ka}\epsilon_{abc}(\tilde Rp)_{bc}.
\end{eqnarray}
To find the Hamiltonian, we exclude the velocities from the expression $H=p_{ij}\dot R_{ij}-L+v_{ij}P_{ij}+\varphi_{ij}\pi_{ij}$, obtaining
\begin{eqnarray}\label{11.6}
H=\frac12 I^{-1}_{ka}[\epsilon_{kij}(\tilde Rp)_{ij}][\epsilon_{abc}(\tilde Rp)_{bc}]+\frac12\lambda_{ij}[R_{ki}R_{kj}-
\delta_{ij}]+v_{ij}P_{ij}+\varphi_{ij}\pi_{ij}.  
\end{eqnarray}
Let us introduce the phase-space functions
\begin{eqnarray}\label{11.7}
M_k(R, p)=-\epsilon_{kij}(\tilde R p)_{ij}. 
\end{eqnarray}
Comparing this expression with Eq. (\ref{11.5}), we see that  $M_k(R, p)$ is the Hamiltonian counterpart of angular momentum in the body, so we denoted it by the same letter. With this notation, we obtain
\begin{eqnarray}\label{11.8}
H=\frac12 I^{-1}_{ij}M_i M_j+\frac12\lambda_{ij}[R_{ki}R_{kj}-\delta_{ij}]+v_{ij}P_{ij}+\varphi_{ij}\pi_{ij}.  
\end{eqnarray}
Now let's move from the canonical variables to the following set:
\begin{eqnarray}\label{11.9}
(R_{ij}, p_{ij}) \rightarrow (R_{ij}, M_i, P_{ij}).
\end{eqnarray}
By construction, it is an invertible (but noncanonical) change of variables in the phase space. 
The Hamiltonian (\ref{11.8}) is already represented through the new variables. Using the canonical brackets $\{ R_{ij}, p_{ij}\}=1$, we get the Poisson brackets of the new variables
\begin{eqnarray}\label{11.20}
\{ R_{ij}, R_{ab}\}=0, \qquad \{M_i, M_j\}=-\epsilon_{ijk}(\tilde R\tilde R^T{\bf M})_k, \qquad 
\{M_i, R_{jk}\}=-\epsilon_{ikm}\tilde R^T_{jm};  
\end{eqnarray}
\begin{eqnarray}\label{11.21}
\{ R_{ij}, P_{ab}\}=\tilde R^T_{ia}\delta_{jb}+\tilde R^T_{ib}\delta_{ja}, \qquad 
\{ M_k, P_{ab}\}=-2M_k(\tilde R\tilde R^T)_{ab}+\delta_{ka}(\tilde R\tilde R^T{\bf M})_b+\delta_{kb}(\tilde R\tilde R^T{\bf M})_a, \cr 
\{ P_{ij}, P_{ab}\}=-(\tilde R\tilde R^T)_{ia}\epsilon_{jbn} M_n-(\tilde R\tilde R^T)_{jb}\epsilon_{jan} M_n+(a\leftrightarrow b).  \qquad \qquad \qquad \qquad 
\end{eqnarray}
For the latter use, we note that  brackets of basic variables $M_a$ and $R_{ab}$ with the orthogonality constraint vanish
\begin{eqnarray}\label{11.22}
\{ M_a, R_{ki}R_{kj}-\delta_{ij} \}=\{ R_{ab}, R_{ki}R_{kj}-\delta_{ij} \}=0.
\end{eqnarray}
Using the obtained brackets, let us discuss the higher-stage constraints. The equation $\dot\pi_{ij}=\{ \pi_{ij}, H\}=0$ implies the orthogonality constraint. In turn, the equation $d(R_{ki}R_{kj}-\delta_{ij})/dt=0$ determines the multipliers: $v_{ij}=0$. At last, as in previous section, the equation $\dot P_{ij}=0$ allows to determine the auxiliary variables $\lambda$ and $\varphi$, but we won't need them. According to Dirac, once all the auxiliary variables have been determined, the final form of the theory is obtained by constructing the Dirac bracket.  Let us denote the twelve independent constraints among $R_{ki}R_{kj}-\delta_{ij}=0$ and $P_{ij}=0$ by $T_A$ and $P_A$, $A=1, 2, \ldots , 6$, and construct the corresponding Dirac bracket
\begin{eqnarray}\label{11.23}
\{ A, B\}_D=\{A, B\}-\{A, \Phi^\alpha\}\triangle^{-1}_{\alpha\beta}\{\Phi^\beta, B\}.
\end{eqnarray}
Here $\Phi^\alpha$ is the set of all constraints: $\Phi^\alpha=(T_A, P_B)$. Besides, denoting symbolically  $6\times 6$\,-blocks as $b=\{ T, P \}$ and $c=\{ P, P \}$, the matrices $\triangle$ and $\triangle^{-1}$ are   
\begin{eqnarray}\label{11.24}
\triangle=\left(
\begin{array}{cc}
0 & b \\
-b^T& c 
\end{array}\right), \qquad 
\triangle^{-1}=\left(
\begin{array}{cc}
b^{-1 T}cb^{-1} & -b^{-1 T} \\
b^{-1} & 0 
\end{array}\right).
\end{eqnarray}
This implies the following structure of the Dirac bracket
\begin{eqnarray}\label{11.25}
\{ A, B\}_D=\{A, B\}-\{A, T\}\triangle ' \{T, B\}+\{A, T\}\triangle '' \{P, B\}.
\end{eqnarray}
Taking into account Eqs. (\ref{11.22}), we conclude that in the passage from Poisson bracket (\ref{11.20}), (\ref{11.21}) to the Dirac bracket, the brackets (\ref{11.20}) of the basic variables will not be modified, retaining their original form. So, fortunately, we do not need to calculate the explicit form of $12\times 12$\,-matrix $\triangle^{-1}$. 

The formulation of the theory in terms of Dirac bracket makes it much more transparent. Indeed, according to the Dirac's formalism, we now can ignore all terms with constraints in the Hamiltonian. Besides, we can use the constraints, in particular $R^TR=1$, before the calculation of the brackets. Doing this in Eqs. (\ref{11.8}) and  (\ref{11.20}) we get
\begin{eqnarray}\label{11.26}
H_0=\frac12 I^{-1}_{ij}M_i M_j;   
\end{eqnarray}
\begin{eqnarray}
\{ R_{ij}, R_{ab}\}_D=0, \qquad \{M_i, M_j\}_D=-\epsilon_{ijk}M_k, \label{11.27} \\
\{M_i, R_{jk}\}_D=-\epsilon_{ikm}R_{jm}.  \label{11.270}
\end{eqnarray}
Denoting the rows of the matrix $R_{ij}$ by ${\bf a}$, ${\bf b}$ and ${\bf c}$, the bracket(\ref{11.270}) read: $\{M_i, a_j\}_D=-\epsilon_{ijk}a_k$,  $\{M_i, b_j\}_D=-\epsilon_{ijk}b_k$, $\{M_i, c_j\}_D=-\epsilon_{ijk}c_k$. Therefore the Poisson structure of a rigid body can be identified with semidirect sum of the algebra $\bar{so}(3)$ with three translation algebras. The constraint's functions $R^TR-{\bf 1}$ and $P_{ij}$ are Casimir functions of the Dirac bracket (\ref{11.23}). 

The brackets (\ref{11.27}) and (\ref{11.270}) were suggested by Chetaev \cite{Chet_1941} as the posible Poisson structure corresponding to the Euler-Poisson equations. We demonstrated, that they can be obtained using the formalism of constrained systems, and are the Dirac brackets that take into account  twelve second-class constraints presented in the theory (\ref{2.1}).

{\bf Integrability of the free equations.} As an application of the Hamiltonian formulation, we present general solution to the Euler-Poisson equations. 
They are obtained according to the rule: $\dot z=\{ z, H_0 \}_D$, and read 
\begin{eqnarray}\label{p26} 
\dot R_{ij}=-\epsilon_{jkm}(I^{-1}M)_k R_{im}, 
\end{eqnarray}
\begin{eqnarray}\label{p27}
\dot{\bf M}=[{\bf M}, I^{-1}{\bf M}].   
\end{eqnarray}
Using the rows ${\bf a}$, ${\bf b}$ and ${\bf c}$, Eqs. (\ref{p26}) can be separated: $\dot{\bf a}=[{\bf a}, I^{-1}{\bf M}]$, 
$\dot{\bf b}=[{\bf b}, I^{-1}{\bf M}]$  and $\dot{\bf c}=[{\bf c}, I^{-1}{\bf M}]$. Then the entire system  (\ref{p26}), (\ref{p27}) breaks down into three. For instance, in the case of the row ${\bf a}$ we have
\begin{eqnarray}\label{p28} 
\dot a_i=[{\bf a}, I^{-1}{\bf M}]_i, \qquad \dot M_j=[{\bf M}, I^{-1}{\bf M}]_j,  
\end{eqnarray}
with the brackets (\ref{11.27}) and $\{M_i, a_j\}_D=-\epsilon_{ijk}a_k$. For the unconstrained Hamiltonian system (\ref{p28}), we can use the known formula of Hamiltonian mechanics to write its solutions  through exponential of the Hamiltonian vector field, see Sect. 2.3 in \cite{deriglazov2010classical}. Applying this formula  we get the solution 
\begin{eqnarray}\label{p29}
M_i(t, M_{0k})=e^{t[{\bf M}_0, I^{-1}{\bf M}_0]_j\frac{\partial}{\partial M_{0j}}}M_{0i}, \quad 
a_i(t, M_{0k})=e^{t([{\bf M}_0, I^{-1}{\bf M}_0]_j\frac{\partial}{\partial M_{0j}}+[{\bf a}_0, I^{-1}{\bf M}_0]_j\frac{\partial}{\partial a_{0j}})}a_{0i}, \cr 
b_i(t, M_{0k})=e^{t([{\bf M}_0, I^{-1}{\bf M}_0]_j\frac{\partial}{\partial M_{0j}}+[{\bf b}_0, I^{-1}{\bf M}_0]_j\frac{\partial}{\partial b_{0j}})}b_{0i}, \quad 
c_i(t, M_{0k})=e^{t([{\bf M}_0, I^{-1}{\bf M}_0]_j\frac{\partial}{\partial M_{0j}}+[{\bf c}_0, I^{-1}{\bf M}_0]_j\frac{\partial}{\partial c_{0j}})}c_{0i}. \quad 
\end{eqnarray}
After computing all derivatives, one should take $a_{0i}=(1, 0, 0)$, $b_{0i}=(0, 1, 0)$ and $c_{0i}=(0, 0, 1)$ in accordance with the initial conditions. The solution depends on three arbitrary constants $M_{0j}$,  and is therefore a general solution to the Euler-Poisson equations. 

In the calculations made in this section were used some specific properties of the group $SO(3)$. However, this formalism can be generalized to the case of a motion on an arbitrary surface, see \cite{AAD23_2}.


\section{Conclusion.}

We have described the theory of a free rigid body using the variational problem (\ref{1.6}) as the sole starting point.
Having accepted the expression (\ref{1.6}), we no longer need any additional postulates or assumptions about the behavior of the rigid body. As shown above, all the basic quantities and characteristics of a rigid body (center of mass, translational and rotational degrees of freedom, mass matrix and tensor of inertia, angular velocity, and so on), as well as the equations of motion and integrals of motion, are obtained from the variational problem by direct and unequivocal calculations within the framework of standard methods of classical mechanics.

From the variational problem we deduced various equivalent system of equations, which can be used to describe the time evolution of rotational degrees of freedom of a free rigid body. All the equations are written in the center-of-mass coordinate system.  They are: \par

\noindent {\bf 1.} The second-order Lagrangian equations (\ref{2.6}) for the rotational degrees of freedom $R_{ij}(t)$. 

\noindent {\bf 2.} The first-order Hamiltonian equations (\ref{2.13})-(\ref{2.15}) for the phase-space degrees of freedom $R_{ij}(t)$ and $\Omega_i(t)$.  

\noindent {\bf 3.} The first-order equations (\ref{6.11}) and (\ref{6.13}) for $R_{ij}(t)$ (which involve three integration constants $m_i$). They were obtained from the Hamiltonian equations of Item 2 excluding $\Omega_i$ with use of the law of preservation of angular momentum of the body. Being rewritten for the Euler angles, they imply Eqs. (\ref{inc.1.6})-(\ref{inc.1.8}). 

Using our formalism, we revisited some cases when solution to the equations of a free rigid body can be obtained in elementary functions.  The solutions (\ref{6.13.05}) and (\ref{s14}) have been found without use of Euler angles. Formulas (\ref{2.15}) and (\ref{s12.11}) solves the problem of motion of the free Lagrange (symmetric) top.

For the free asymmetric body, its general solution can be written through exponential of the Hamiltonian vector field, see Eq. (\ref{p29}). 

In conclusion, we list some noticed in this work features, that are not taken into account in standard textbooks when formulating the laws of motion of a rigid body. These peculiar properties are a consequence of the fact that not all solutions to the equations of motion of a rigid body describe the motions of the body.

\noindent {\bf A.} By costruction of the variables $R_{ij}$ (see Eq. (\ref{1.19})), their equations of motion should be supplemented by the universal initial condition $R_{ij}(0)=\delta_{ij}$, that is we are interested only in the trajectories which pass through unit 
element of $SO(3)$. Only these solutions correspond to the movements of a rigid body. When we work with a rigid body in terms of Euler angles, this implies that $\theta(t)\rightarrow 0$ as $t\rightarrow 0$. 

\noindent {\bf B.} According to Eq. (\ref{4.3.0}), the initial conditions for angular velocity are fixed by the conserved angular momentum. 

\noindent {\bf C.} According to Eq. (\ref{2.8.1}), the rotational energy of a rigid body does not represent an independent integral of motion. 

\noindent {\bf D.}  For the convenience of calculations, all three systems of equations are written with a diagonal inertia tensor. This implies certain restrictions on the range of applicability of these equations: ${\bf R}_i(0)={\bf b}_i(0)={\bf e}_i$. That is, at the initial moment of time, the body fixed basis and the axes of inertia coincide with the basis vectors of the Laboratory. For the case of asymmetric body ($I_1\ne I_2\ne I_3$) this implies, in particular, that we cannot perform the rotation of the Laboratory system in order to simplify the equations of motion.

\begin{acknowledgments}
The work has been supported by the Brazilian foundation CNPq (Conselho Nacional de
Desenvolvimento Cient\'ifico e Tecnol\'ogico - Brasil). 
\end{acknowledgments}

\section*{Appendix 1. Equations of motion in terms of Euler angles.}\label{Ap1}

The equations of motion of a rigid body, discussed in Sect. \ref{EA}, are still written for an excess number of variables. Indeed, at each instant of time, the nine matrix elements $R_{ij}$ obey to six constraints $R^TR =1$, so we need to know only some $9-6=3$ independent parameters to specify the matrix $R$. 
There are many different ways to parameterize the rotation matrices \cite{Lei_1965,Gol_2000,Arn_2,AAD23_1}.
We discuss one possible choice for these parameters, called the Euler angles. Their geometric meaning is as follows: according to Eq. (\ref{1.23}), an orthogonal matrix $R$ can be considered as composed of three orthonormal vectors ${\bf R}_i$ rigidly connected with the body. This basis can be obtained from the Laboratory 
basis ${\bf e}_i$ by making of a sequence of three rotations about suitably chosen axes. These three rotation angles are just the Euler angles. They can be defined according to the following rule. If the vector ${\bf R}_3$ is not collinear with ${\bf e}_3$, we calculate the vector product $[{\bf e}_3, {\bf R}_3]\equiv {\bf e}'_1$, and construct an intermediate coordinate axis in the direction of the vector ${\bf e}'_1$. Now, to turn out the basis ${\bf e}_i$ into ${\bf R}_i$, we can do the following sequence of rotations: \par
\begin{figure}[t] \centering
\includegraphics[width=17cm]{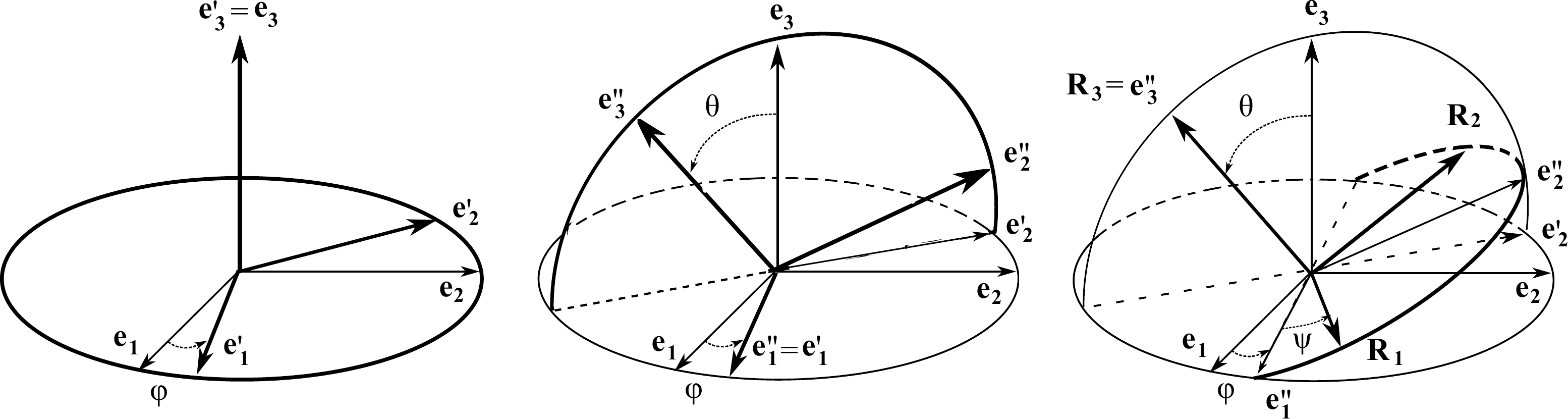}
\caption{Definition of the Euler angles $\varphi$, $\theta$ and $\psi$.}\label{TvT_6}
\end{figure}

 
\noindent 1. At an angle $\varphi$ counterclockwise about the axis\footnote{The rotation should be counterclockwise when viewed from the end of the rotation vector.} ${\bf e}_3$ so that ${\bf e}_1$ turn into ${\bf e}'_1$. This turn out three basic vectors ${\bf e}_i$ into ${\bf e}'_i$, see Figure \ref{TvT_6}. \par 
\noindent 2. At an angle $\theta$ counterclockwise about the axis ${\bf e}'_1$ so that ${\bf e}_3$ turn into ${\bf R}_3\equiv{\bf e}''_3$. This gives three  vectors ${\bf e}''_i$. \par 
\noindent 3. At an angle $\psi$ counterclockwise about the axis ${\bf e}''_3={\bf R}_3$ so that ${\bf e}''_1={\bf e}'_1$ turn into ${\bf R}_1$ and ${\bf e}''_2$ into ${\bf R}_2$. Then the resulting vectors are ${\bf R}_i$. 

In the region 
\begin{eqnarray}\label{inc.1}
0<\varphi<2\pi,  \qquad 0<\theta<\pi, \qquad 0<\psi<2\pi, 
\end{eqnarray}
the constructed map $(\varphi, \theta, \psi) \rightarrow R_{ij}$ determines local coordinates on the surface $R^TR =1$. The angle $\theta$ has a simple visualization, being the angle that determinis the cone with axis ${\bf e}_3$, on which lies the body-fixed vector ${\bf R}_3$. It is convenient to introduce the angle $\alpha\equiv (3\pi/2+\varphi)/mod ~ 2\pi$. Then $\theta, \alpha$ are spherical coordinates (altitude and azimuth) of the body-fixed vector ${\bf R}_3$. 

Let us present manifest expressions of these rotations, as well as the inverse transformations and the variation rates of basic vectors with time. \par 
\noindent The rotation $\varphi$: 
\begin{eqnarray}
{\bf e}'_1={\bf e}_1\cos\varphi+ {\bf e}_2\sin\varphi, \quad {\bf e}'_2=-{\bf e}_1\sin\varphi+ {\bf e}_2\cos\varphi, \quad {\bf e}'_3={\bf e}_3; \label{6.1.1} \\
{\bf e}_1={\bf e}'_1\cos\varphi-{\bf e}'_2\sin\varphi, \quad {\bf e}_2={\bf e}'_1\sin\varphi+{\bf e}'_2\cos\varphi; \label{6.1.2} \\
\dot{\bf e}'_1=\dot\varphi{\bf e}'_2, \quad \dot{\bf e}'_2=-\dot\varphi{\bf e}'_1, \quad \dot{\bf e}'_3=0. \label{6.1.3}
\end{eqnarray}
\noindent The rotation $\theta$: 
\begin{eqnarray}
{\bf e}''_1={\bf e}'_1,  \quad {\bf e}''_2={\bf e}'_2\cos\theta+ {\bf e}'_3\sin\theta, \quad {\bf e}''_3=-{\bf e}'_2\sin\theta+{\bf e}'_3\cos\theta; \label{6.2.1} \\
{\bf e}'_2={\bf e}''_2\cos\theta-{\bf e}''_3\sin\theta, \quad {\bf e}'_3={\bf e}''_2\sin\theta+{\bf e}''_3\cos\theta; \label{6.2.2} \\
\dot{\bf e}''_1=\dot{\bf e}'_1=\dot\varphi{\bf e}'_2, \quad \dot{\bf e}''_2=-{\bf e}'_1\dot\varphi\cos\theta-{\bf e}'_2\dot\theta\sin\theta+
{\bf e}'_3\dot\theta\cos\theta,  
\quad \dot{\bf e}''_3={\bf e}'_1\dot\varphi\sin\theta-{\bf e}'_2\dot\theta\cos\theta-{\bf e}'_3\dot\theta\sin\theta. \label{6.2.3}
\end{eqnarray}
\noindent The rotation $\psi$: 
\begin{eqnarray}
{\bf R}_1={\bf e}''_1\cos\psi+{\bf e}''_2\sin\psi, \quad {\bf R}_2=-{\bf e}''_1\sin\psi+{\bf e}''_2\cos\psi, \quad {\bf R}_3={\bf e}''_3; \label{6.3.1} \\
{\bf e}''_1={\bf R}_1\cos\psi-{\bf R}_2\sin\psi, \quad {\bf e}''_2={\bf R}_1\sin\psi+{\bf R}_2\cos\psi.  \label{6.3.2} \\
\end{eqnarray}
Using Eqs. (\ref{6.2.1}) and (\ref{6.1.1}) in (\ref{6.3.1}), we can present the rotated vectors ${\bf R}_i$ through the initial ${\bf e}_i$. This gives elements of the rotation matrix (\ref{1.23}) in terms of the Euler angles: 
\begin{eqnarray}\label{6.3}
R=\left(
\begin{array}{ccc}
\cos\psi\cos\varphi-\sin\psi\cos\theta\sin\varphi  &  -\sin\psi\cos\varphi-\cos\psi\cos\theta\sin\varphi &\sin\theta\sin\varphi \\
\cos\psi\sin\varphi+\sin\psi\cos\theta\cos\varphi & -\sin\psi\sin\varphi+\cos\psi\cos\theta\cos\varphi & -\sin\theta\cos\varphi \\
\sin\psi\sin\theta & \cos\psi\sin\theta & \cos\theta
\end{array}\right)
\end{eqnarray}
Computing derivatives of columns, we get the variation rates of moving basis vectors as follows:
\begin{eqnarray}\label{6.4}
\dot{\bf R}_1=(\dot\psi+\dot\varphi\cos\theta){\bf R}_2+(-\dot\varphi\cos\psi\sin\theta+\dot\theta\sin\psi){\bf R}_3,  \cr
\dot{\bf R}_2=-(\dot\psi+\dot\varphi\cos\theta){\bf R}_1+(\dot\varphi\sin\psi\sin\theta+\dot\theta\cos\psi){\bf R}_3, \cr
\dot{\bf R}_3=(\dot\varphi\cos\psi\sin\theta-\dot\theta\sin\psi){\bf R}_1+(-\dot\varphi\sin\psi\sin\theta-\dot\theta\cos\psi){\bf R}_2. 
\end{eqnarray}
Computing $\frac12[{\bf R}_i, \dot{\bf R}_i]$ and representing the result in the form ${\bf e}_j\omega_j$, we obtain components of angular velocity
\begin{eqnarray}\label{6.5}
\omega_j=(\dot\theta\cos\varphi+\dot\psi\sin\theta\sin\varphi, \quad \dot\theta\sin\varphi-\dot\psi\sin\theta\cos\varphi, \quad \dot\varphi+\dot\psi\cos\theta),
\end{eqnarray}
while representing the result in the form ${\bf R}_j\Omega_j$, we obtain angular velocity in the body
\begin{eqnarray}\label{6.6}
\Omega_j=(\dot\varphi\sin\psi\sin\theta+\dot\theta\cos\psi, \quad \dot\varphi\cos\psi\sin\theta-\dot\theta\sin\psi, \quad \dot\psi+\dot\varphi\cos\theta). 
\end{eqnarray}

Substituting $R_{ij}(\varphi, \theta, \psi)$ of Eq. (\ref{6.3}) into the equations of motion (\ref{6.11}) and (\ref{6.12}),  we obtain a system of ten equations for determining the temporal evolution of Euler angles. We can separate any three of equations which unambiguoisly fix these three angles, and then try to solve them. According to (\ref{6.13}), the equation (\ref{6.12}) in terms of Euler angles is the algebraic equation. So one of the Euler angles can be found from this equation in terms of the other two angles. These last should be found by solving two differential equations separated from the system (\ref{6.11}).

Here it is instructive to discuss some peculiarities of the Euler coordinate system.

\noindent 
{\bf 1.} The intermediate axis is not defined when ${\bf R}_3={\bf e}_3$. Due to this, even in small vicinity of unit matrix $R_{ij}=\delta_{ij}$, not all matrices acquire the Euler coordinates. They are the matrices that correspond to rotations on a small angle $\beta$ around the axis ${\bf e}_3$. In particular, the Euler angles are not defined for the unit matrix $\delta_{ij}$. Therefore, having written our  equations (\ref{6.11}) in terms of Euler angles, we cannot write down the initial data for them that would correspond to the condition $R_{ij}(t_0)=\delta_{ij}$. We recall that only such a kind solutions of the equations (\ref{6.11})-(\ref{6.10}) describe the motions of the rigid body. Therefore, having obtained some solution $\varphi(t), \theta(t), \psi(t)$ in the region (\ref{inc.1}), we must check that in the limit $t\rightarrow t_0$ for some $t_0$ the matrix $R_{ij}(\varphi(t), \theta(t), \psi(t))$ approaches the identity matrix. For the Euler angles in this limit we must get 
\begin{eqnarray}\label{inc.1.0}
\theta(t)\rightarrow 0, \quad \varphi(t)+\psi(t)\rightarrow 0 ~ \mbox{or} ~   2\pi \quad \mbox{as} \quad t\rightarrow t_0.
\end{eqnarray}
Indeed, $\theta(t)$ should approach zero by construction of this coordinate. For the small $\theta$, using $\cos\theta\sim 1-\frac12\theta^2$ and $\sin\theta\sim\theta$ in Eq. (\ref{6.3}), we get
\begin{eqnarray}\label{6.30}
R\sim \left(
\begin{array}{ccc}
\cos(\varphi+\psi) &  -\sin(\varphi+\psi) & 0  \\
\sin(\varphi+\psi) & \cos(\varphi+\psi) &  0 \\
0 & 0 & 1
\end{array}\right)+ O(\theta). 
\end{eqnarray}
So, the condition $R_{ij}\rightarrow\delta_{ij}$ implies $\varphi(t)+\psi(t)\rightarrow 0, 2\pi$ as $t\rightarrow t_0$.

In Sect. \ref{IN} we obtained an example of the solution which does not obey these conditions. Indeed, for the the solution (\ref{6.13.05}) we 
have $\theta(t)=0$, that is it lies outside the Euler coordinate system for all $t$. 

In this regard, we note that matrices close to identity do not necessarily have small coordinates. For instance, consider the matrix that corresponds to the rotation on small angle $t$ counterclockwise around the axis ${\bf e}_1$. Its Euler coordinates are $\varphi=0, \theta=t, \psi=0$. But if we do the clockwise rotation on small angle $\tau$ around ${\bf e}_1$, the Equler coordinates 
are $\varphi=\pi, \theta=\tau, \psi=\pi$. The curves $R_{ij}(t)$ and $R_{ij}(\tau)$ approach $\delta_{ij}$ when their parameters approach to $0$. 

\noindent 
{\bf 2.} Substituting the matrix $R_{ij}$ taken in the Euler parametrization (\ref{6.3}), into our equations (\ref{6.11}), we obtain nine equations of the form
\begin{eqnarray}\label{inc.1.1}
\frac{\partial R_A}{\partial\alpha_k}\dot\alpha_k=f_A(\alpha_k), 
\end{eqnarray}
where $\alpha_1=\varphi, \alpha_2=\theta, \alpha_3=\psi$, and $R_A$ is the column $({\bf R}_1, {\bf R}_2, {\bf R}_3)^T$. In a vicinity of any point $(\varphi, \theta\ne 0, \psi)$ we get $rank \frac{\partial R_A}{\partial\alpha_k}=3$, so we can separate three equations of the system (\ref{inc.1.1}), 
which further can be presented in the normal form $\dot\alpha_i=f_i(\alpha_k)$. However, note that for the values $\alpha_i=0$, rank of this matrix is equal to $2$. 

Let us present manifest form for the following equations of the system (\ref{6.11}): 
\begin{eqnarray}\label{inc.1.2}
\dot R_{33}=-\Omega_1R_{32}+\Omega_2R_{31}, \quad  \dot R_{31}=-\Omega_2R_{33}+\Omega_3R_{32}, \quad \dot R_{13}=-\Omega_1R_{12}+\Omega_2R_{11}.
\end{eqnarray}
Substituting (\ref{6.3}), they can be written as follows:
\begin{eqnarray}
\dot\theta\sin\theta=(\Omega_1\cos\psi-\Omega_2\sin\psi)\sin\theta, \label{inc.2.1} \\
\dot\varphi\sin\theta=\Omega_1\sin\psi+\Omega_2\cos\psi, \label{inc.2.3} \\
\dot\psi\sin\theta=-(\Omega_1\sin\psi+\Omega_2\cos\psi)\cos\theta+\Omega_3\sin\theta. \label{inc.2.2}
\end{eqnarray}
These equations follow also from Eq. (\ref{6.6}).
The explicit expressions for the components of angular velocity (\ref{6.13}) are:
\begin{eqnarray}
\Omega_1=\frac{m_1}{I_1}[\cos\psi\cos\varphi-\sin\psi\cos\theta\sin\varphi]+\frac{m_2}{I_1}[\cos\psi\sin\varphi+\sin\psi\cos\theta\cos\varphi]+\frac{m_3}{I_1}\sin\psi\sin\theta, \label{inc.1.3} \\
\Omega_2=\frac{m_1}{I_2}[-\sin\psi\cos\varphi-\cos\psi\cos\theta\sin\varphi]+\frac{m_2}{I_2}[-\sin\psi\sin\varphi+\cos\psi\cos\theta\cos\varphi]+\frac{m_3}{I_2}\cos\psi\sin\theta, \label{inc.1.4} \\ 
\Omega_3=\frac{m_1}{I_3}\sin\theta\sin\varphi-\frac{m_2}{I_3}\sin\theta\cos\varphi+\frac{m_3}{I_3}\cos\theta. \label{inc.1.5}
\end{eqnarray}
Substituting  them into the previous equations we get 
\begin{eqnarray}
\sin\theta[\dot\theta-(m_1\cos\varphi+m_2\sin\varphi)(\frac{1}{I_1}\cos^2\psi+\frac{1}{I_2}\sin^2\psi)- I_{(1-2)}[(m_2\cos\varphi-m_1\sin\varphi)\cos\theta+m_3\sin\theta]\sin\psi\cos\psi]=0, \quad \label{inc.1.6} \\
\dot\varphi\sin\theta=I_{(1-2)}(m_1\cos\varphi+m_2\sin\varphi)\sin\psi\cos\psi+(\frac{1}{I_1}\sin^2\psi+\frac{1}{I_2}\cos^2\psi)[(m_2\cos\varphi-m_1\sin\varphi)\cos\theta+m_3\sin\theta], \quad \label{inc.1.7} \\
\dot\psi\sin\theta=-[\dot\varphi\sin\theta]\cos\theta+\frac{1}{I_3}[-(m_2\cos\varphi-m_1\sin\varphi)\sin^2\theta+m_3\sin\theta\cos\theta], \quad \label{inc.1.8}
\end{eqnarray}
where it was denoted $I_{(k-p)}=\frac{1}{I_k}-\frac{1}{I_p}$. These equations together with (\ref{6.12}) can be thought as equations of motion of a rigid body in terms of Euler angles. 


{\bf Discussion of the solution (\ref{6.13.05}).} Let us try to use the Euler coordinates to reproduce the solution (\ref{6.13.05}), describing the motion of asymmetrical body with special initial conditions ${\bf m}=(0, 0, m_3)$. 
In terms of Euler angles, this solution should have the following structure: $(\varphi(t), \theta(t)=0, \psi(t))$, that is, it lies outside the Euler coordinate system (\ref{inc.1}). So one cannot expect that the solution could be found by solving the system in terms of Euler angles. 
Let's see what happens, if we nevertheless try to do this, taking the conserved angular momentum equal to (\ref{in.1}). Substitution of (\ref{in.1}) into (\ref{inc.1.6})-(\ref{inc.1.8}) gives the equations 
\begin{eqnarray}\label{in.2}
\dot\theta=m_3 I_{(1-2)}\sin\theta\sin\psi\cos\psi, \cr
\dot\varphi=\frac{m_3}{I_1}\sin^2\psi+\frac{m_3}{I_2}\cos^2\psi,  \cr
\dot\psi=-(\frac{m_3}{I_1}\sin^2\psi+\frac{m_3}{I_2}\cos^2\psi)\cos\theta+\frac{m_3}{I_3}\cos\theta. 
\end{eqnarray}
We point out that these equations coincide with those written on page 145 of the book \cite{Whit_1917}. So our results, presented below, should be compared with those of Sect. 69 in \cite{Whit_1917}, as well as of Sect. 88-93 in \cite{Mac_1936}. 

We start our analysis from the algebraic equation (\ref{6.12}), that acquires the form 
\begin{eqnarray}\label{6.14.1}
2E=\sum_i I_i\Omega_i^2=m_3^2(\frac{1}{I_1}\sin^2\psi\sin^2\theta+\frac{1}{I_2}\cos^2\psi\sin^2\theta+\frac{1}{I_3}\cos^2\theta)= \cr \frac{m^2_3}{I_3}+m^2_3[I_{(1-3)}-I_{(1-2)}\cos^2\psi]\sin^2\theta.
\end{eqnarray}
Remarkably, this expression does not involve $\varphi$ at all. If the solution we are looking for describes the motion of a rigid body, it must satisfy the relation (\ref{2.8.1}) between the integration constants $E$ and $m_i$. Using (\ref{2.8.1})  in (\ref{6.14.1}), we obtain the following equation: 
\begin{eqnarray}\label{6.14.2}
[I_{(1-3)}-I_{(1-2)}\cos^2\psi]\sin^2\theta=0.
\end{eqnarray}
Since we work in vicinity of a point with $\theta\ne 0$, this equation implies that the angle $\psi$ does not change with time
\begin{eqnarray}\label{6.14.3}
\psi=\psi_0, \quad \mbox{such that} \quad \cos^2\psi_0=\frac{I_{(1-3)}}{I_{(1-2)}}=\frac{I_2(I_3-I_1)}{I_3(I_2-I_1)}=\frac{g_1^2-g_3^2}{g_1^2-g_2^2}.
\end{eqnarray}
The last equation of the system (\ref{in.2}) is satisfied by $\psi(t)=\psi_0$, while the remaining two equations read
\begin{eqnarray}\label{6.14.4}
\dot\theta=\frac12 m_3I_{(1-2)}\sin2\psi_0 \sin\theta, \qquad 
\dot\varphi=\frac{m_3}{I_3}, 
\end{eqnarray}
and can be immediately integrated. In the result, we get the following solution to the rigid body equations of motion 
\begin{eqnarray}
t=\frac{2}{m_3I_{(1-2)}\sin2\psi_0}\int\frac{d\theta}{\sin\theta}+c', \label{6.14.6} \\
\varphi=\frac{m_3}{I_3}t+\varphi_0,  \label{6.14.7} \\
\psi=\psi_0. \label{6.14.8}
\end{eqnarray}
Denote $\frac12 m_3I_{(1-2)}\sin2\psi_0\equiv k$. We have $k=0$ when $\psi_0=0, \pi/2, \pi$ and  $3\pi/2$. But $\cos^2\psi_0=I_{(1-3)}/I_{(1-2)}$, so we can assume $k\ne 0$. Computing integral in (\ref{6.14.6}) we get
\begin{eqnarray}\label{6.14.9}
\cos\theta(t)=\frac{1-ce^{2kt}}{1+ce^{2kt}}, \qquad \mbox{where} \quad  c>0. 
\end{eqnarray}
We are interested in a solution with the property $\lim_{t\rightarrow t_0}\theta(t)=0$ for some finite value $t_0$. For any value of the integration constant $c$, there is no such $t_0$.  So none of the solutions (\ref{6.14.6})-(\ref{6.14.8}) describes the motion of a rigid body.

\section*{Appendix 2. Equivalence of two Lagrangians.}\label{Ap2}
In the Lagrangian  
\begin{eqnarray}\label{ap.1}
L=\frac12 g_{ij}\dot R_{ki}\dot R_{kj} -\frac12 \lambda_{ij}\left[R_{ki}R_{kj}-\delta_{ij}\right],
\end{eqnarray}
the matrix $R_{ij}$ should be treated as an arbitrary matrix (not orthogonal). Due to this, the equivalence of this theory with (\ref{11.2}) is not obvious, and requires confirmation. Writting (\ref{ap.1}) in terms of $M_i(R, \dot R)=-\frac12 I_{ik}\epsilon_{knm}(R^T\dot R)_{nm}$ we get  
\begin{eqnarray}\label{ap.2}
L=\frac12 I^{-1}_{ij}M_i M_j-\frac12\lambda_{ij}[(R^TR)_{ij}-\delta_{ij}]-\frac12 G_{ij}(R, \dot R)[(R^TR)^{-1}_{ij}-\delta_{ij}],
\end{eqnarray}
where $G_{ij}\equiv \hat\Omega_{ia}g_{ab}\hat\Omega_{bj}$. This theory is equivalent to 
\begin{eqnarray}\label{ap.3}
L_1=\frac12 I^{-1}_{ij}M_i M_j-\frac12[\lambda_{ij}+G_{ij}][R_{ki}R_{kj}-\delta_{ij}],
\end{eqnarray}
this can be easily verified by comparing their equations of motion. Now, consider the theory (\ref{11.2}): 
\begin{eqnarray}\label{ap.4}
L_2=\frac12 I^{-1}_{ij}M_i M_j-\frac12\lambda '_{ij}[R_{ki}R_{kj}-\delta_{ij}], 
\end{eqnarray}
and compare the equations of motion of $L_1$ and $L_2$. For $L_2$ they have the structure
\begin{eqnarray}\label{ap.5}
D_{ij}(\ddot R, \dot R, R)+\lambda '_{ik}R_{kj}=0, \qquad R^TR=1, 
\end{eqnarray}
while for $L_1$ we get
\begin{eqnarray}\label{ap.6}
D_{ij}(\ddot R, \dot R, R)+[\lambda_{ik}+G_{ik}]R_{kj}=0, \qquad R^TR=1, 
\end{eqnarray}
with the same $D_{ij}$. 
If $R_0(t)$, $\lambda '_0(t)$ is a solution to (\ref{ap.5}), then $R_0(t)$, $\lambda_0(t)=\lambda '_0(t)-G(\dot R_0, R_0)$ is a solution to (\ref{ap.6}).  Conversely, if $R_0(t)$, $\lambda_0(t)$ is a solution to (\ref{ap.6}), then $R_0(t)$, $\lambda '_0(t)=\lambda_0(t)+G(\dot R_0, R_0)$ is a solution to (\ref{ap.5}). The theories (\ref{ap.1}) and (\ref{ap.4}) have the same solutions for the confuguration-space variables $R_{ij}(t)$, and henceforth are equivalent.

\end{document}